\documentclass{jfm}

\usepackage{amsmath, amsfonts, amssymb}
\usepackage{graphicx}
\usepackage{tikz}
\usepackage{color}
\usepackage{float}
\usepackage{subcaption}
\usepackage[natbib=true,style=authoryear]{biblatex}
\usepackage[colorlinks=true, linkcolor=red, anchorcolor=black, citecolor=blue]{hyperref}

\newcommand{\lr}[1]{\left(#1\right)}
\newcommand{\pdeone}[2]{\frac{\partial #1}{\partial #2}}
\newcommand{\pden}[3]{\frac{\partial^{#3} #1}{\partial #2^{#3}}}
\newcommand{\odeone}[2]{\frac{\mathrm{d} #1}{\mathrm{d} #2}}

\newcommand{\bld}[1]{\mathbf{#1}}
\renewcommand{\div}{\nabla\cdot}

\newcommand{\nund}{\nu_{\star}}
\newcommand{\kand}{\kappa_{\star}}

\newcommand{\flowmap}{flow map }
\newcommand{\RB}{Rayleigh-B\'{e}nard }
\newcommand{\OB}{Oberbeck-Boussinesq }

\renewcommand{\ss}[1]{_{_{#1}}} 

\makeatletter
\newcommand*{\rom}[1]{\expandafter\@slowromancap\romannumeral #1@}
\makeatother

\begin{document}

\title[Optimal Heat Transport in \RB Convection]{Optimal Heat Transport Solutions for \RB Convection}

\author[D.Sondak, L.M. Smith and F.Waleffe]%
{D.Sondak$^1$%
  \thanks{Email address for correspondence: sondak@math.wisc.edu},\ns
L.M.Smith$^1$$^{,2}$ and F.Waleffe$^1$$^{,2}$}

\affiliation{$^1$Department of Mathematics, University of Wisconsin-Madison,
Madison, WI 53706 $^{2}$Department of Engineering Physics, University of Wisconsin-Madison,
Madison, WI 53706}

\pubyear{???}
\volume{???}
\pagerange{???--???}
\date{?; revised ?; accepted ?. - To be entered by editorial office}
\maketitle

\begin{abstract}
Steady flows that optimize heat transport are obtained for two-dimensional \RB convection with no-slip horizontal walls for a variety of Prandtl numbers $Pr$ and Rayleigh number up to $Ra\sim 10^{9}$.  Power law scalings of $Nu \sim Ra^{\gamma}$ are observed with $\gamma\approx 0.31$, where the Nusselt number $Nu$ is a non-dimensional measure of the vertical heat transport.  Any dependence of the scaling exponent on $Pr$ is found to be extremely weak.  On the other hand, the presence of two local maxima of $Nu$ with different horizontal wavenumbers at the same $Ra$ leads to the emergence of two different flow structures as candidates for optimizing the heat transport.  For $Pr \lesssim 7$,  optimal transport is achieved at the smaller maximal wavenumber.  In these fluids, the optimal structure is a plume of warm rising fluid which spawns left/right horizontal arms near the top of the channel, leading to downdrafts adjacent to the central updraft.  For $Pr > 7$ at high-enough $Ra$, the optimal structure is a single updraft absent significant horizontal structure, and characterized by the larger maximal wavenumber.
\end{abstract}
\section{Introduction}
Thermal convection is a pervasive phenomenon found in a wide variety of natural and engineering processes~\citep{lappa2009thermal}.  Examples include motion of the Earth's molten core, atmospheric and oceanic dynamics and solar phenomena.  A common approximation in fundamental studies of convection is that all fluid parameters are constant except for the fluid density in the buoyancy force~\citep{lappa2009thermal, chandrasekhar2013hydrodynamic, drazin2004hydrodynamic}.  This is the \OB approximation, valid in many physical situations, that allows systems to be described by the incompressible Navier-Stokes equations subject to a buoyancy force and an advection-diffusion equation for the temperature field.  The canonical problem is \RB convection which focuses on flow fields driven by buoyancy and confined between two infinite horizontal plates.  \RB convection was an early study of hydrodynamic instability in which a given flow configuration is unstable at certain wavenumbers as a control parameter is varied. In this case, the control parameter is the Rayleigh number $Ra$ characterizing the relative strength of buoyancy driven inertial forces to viscous forces. In atmospheric and solar environments, estimates of $Ra$ are typically exceptionally large.  Caution must be exercised in such parameter regimes, however, because the \OB approximation may no longer be valid.

A basic question in thermal convection is how much heat is transported for a given temperature difference.  For the \RB problem this is asking how the non-dimensional heat flux $Nu$ depends on the Rayleigh number $Ra$.  For sufficiently large $Ra$, one expects a power law scaling $Nu\sim Ra^{\gamma}$ characteristic of a statistically self-similar asymptotic regime.  Transitions between power laws would indicate significant changes in the structure of the flow fields transporting the heat.  Rigorous mathematical upper bounds on heat transport provide support for power law scaling $Nu\sim Ra^{\gamma}$ and establish that $\gamma \le 1/2$ for no-slip boundary conditions. Next, we briefly review the theoretical results emanating from upper bound theories and scaling arguments.  We then briefly describe experimental and computational results.  Comprehensive details on the history of \RB convection can be found in various reviews~\citep{siggia1994high, ahlers2009heat, chilla2012new}.

The original idea and the inspiration for determining upper bounds on heat transport is due to \citet{malkus1954heat} who sought to determine maximum heat transport subject to integral constraints together with marginal stability assumptions on the mean temperature profile and the smallest scale of motion.  \citet{howard1963heat} removed the marginal stability assumptions from Malkus' theory to derive mathematically rigorous upper bounds on heat transport. Howard attempted to maximize $Nu$ subject to `power integrals' derived from the \OB equations and the no-slip boundary conditions.  In such upper bound theories, the class of possible optimizing fields is expanded; that is, the solutions to the variational problem are not required to satisfy the full \OB equations.  Howard performed his analysis with and without the continuity constraint.  Without the constraint of continuity he found $Nu\ss{B}\sim \lr{3Ra/64}^{1/2}$ for large $Ra$, where $Nu\ss{B}$ denotes the upper bound on $Nu$.  The incorporation of the continuity equation is more mathematically challenging.  Assuming that the optimum contained only one horizontal wavenumber, Howard derived $Nu\ss{B}\sim \lr{Ra/33}^{3/8}$.  However, \citet{busse1969howard}  showed that multi-wavenumber, divergence-free fields have higher heat transport and he derived $Nu\ss{B}\sim \lr{Ra/1035}^{1/2}$ using an approximate nested boundary layers approach.

In the 1990s, an alternative bounding approach called the background field method was developed~\citep{doering1992energy,doering1994variational, constantin1995variational, doering1996variational}.  The basic idea is to decompose the velocity field into a background field and fluctuations about the background.  The background fields are used as trial functions in the variational statement to maximize desired flow quantities (such as $Nu$).  When applied to the \RB problem with no-slip boundary conditions, the method yields an upper bound of $Nu\ss{B}\sim 0.167Ra^{1/2}-1$.  \citet{kerswell2001new} demonstrated a connection between the Malkus-Howard-Busse method and the Doering-Constantin background field method, illustrating that the two methods approach the same bound, with approach from below (above) in the Malkus-Howard Busse (Doering-Constantin) approach.  The previous upper bounds were improved, and shown to be ``sandwiched'' between $0.031Ra^{1/2}$ (derived from the Malkus-Howard-Busse theory by Busse's approximate nested boundary layers approach ) and $0.0335Ra^{1/2}$ (from the background field method).  Although we focus on the no-slip case in this work, it is important to note that an upper bound for the two-dimensional free-slip case of $\gamma=5/12 < 1/2$ was recently found~\citep{whitehead2011ultimate}.  That result is significant because it indicates that boundary conditions and dimensionality play a role in limiting heat transport.  Recently~\citet{wen2015aspect} sharpened the free-slip bound and derived $Nu\ss{B}-1\lesssim 0.106Ra^{5/12}$.  Extending the results on rigorous upper bounds to include Prandtl number ($Pr = \nu/\kappa$) effects has been challenging although some progress has been made in the infinite $Pr$ limit~\citep{whitehead2014rigorous}.

The works discussed so far focused on developing rigorous upper bounds on heat transport.  There are also scaling results based on classic turbulence modeling.  \citet{kraichnan1962turbulent} used eddy viscosity and mixing length ideas to develop a scaling theory of $Nu$ as a function of $Ra$ and $Pr$.  His work led to a prediction of an ultimate regime with $Nu \sim Ra^{1/2}$ and $Pr$-dependence as follows:  $Nu \sim Pr^{1/2}Ra^{1/2}\left[\ln\lr{Ra}\right]^{-3/2}$ for $Pr \leq 0.15$; $Nu\sim Pr^{-1/4}Ra^{1/2}\left[\ln\lr{Ra}\right]^{-3/2}$ for $0.15 < Pr \leq 1$.  Later, \citet{shraiman1990heat} found $Nu\approx 0.27Pr^{-1/7}Ra^{2/7}$ assuming the existence of turbulent boundary layers.  The more recent Grossmann-Lohse (GL) theory decomposes dissipation rates into bulk and boundary layer contributions~\citep{grossmann2000scaling, grossmann2001thermal, grossmann2002prandtl, grossmann2004fluctuations}.  The Grossmann-Lohse framework results in two equations for $Nu\lr{Ra,Pr}$ and $Re\lr{Ra,Pr}$ with five parameters that are to be fit using experimental data.  The theory has also been adapted to the case of turbulent boundary shear layers~\citep{grossmann2011multiple} yielding $Nu\sim \lr{PrRa}^{1/2}\times \text{logarithmic corrections}$.  Recent updates to the parameters based on experiments with aspect ratio $\Gamma$ less than or equal to one~\citep{stevens2013unifying} have led to excellent agreement of this theory with existing experimental data~\citep{niemela2006turbulent, he2012heat, he2012transition}.  A major contribution of the GL theory has been the conceptual breakdown of the scaling laws into the $Pr-Ra$ phase space.  Our present work corresponds to regions $\rom{1}_{u}$ and $\rom{2}_{u}$ of this phase diagram as well as the beginning of region $\rom{4}_{u}$.  In each of these regions the thermal boundary layer is nested within the viscous boundary layer.

A large number of experiments have been conducted to study \RB convection.  It is not expected that the rigorous theoretical upper bounds described in the beginning of this section will be realized, although some experimental evidence does indicate that the ultimate regime may be attained~\citep{chavanne2001turbulent, he2012transition}.  A main focus of experiments is on determining just how turbulent heat transport behaves in nature, as well as on exploring the limits of the \OB approximation in such systems~\citep{niemela2006turbulent}.  A variety of fluids have been used as working fluids in experiments~\citep[see][for an overview]{chilla2012new}. Rayleigh numbers as high as $Ra \approx 10^{17}$ have been reached using cryogenic helium gas~\citep{niemela2000turbulent}.  In most of these experiments it appears that the scaling of $Nu$ is steeper than $Ra^{2/7}$ and slightly below $Ra^{1/3}$ (with notable exceptions including~\cite{chavanne2001turbulent},~\cite{roche2001observation} and~\citet{he2012transition}).
Indeed,~\citet{niemela2000turbulent} found $Nu\sim Ra^{0.309}$.  We also note that the $Pr$ dependence of these scaling laws is found to be very weak for $Pr\gtrsim 1$~\citep{ahlers2009heat}.  Moreover, recent experimental results have been interpreted as a transition to the ultimate regime of thermal convection~\citep{he2012heat, he2012transition}.  Those results suggest that the classical scaling of $Nu\sim Ra^{1/3}$ persists up to $Ra\approx 10^{13}$ after which a transition to the ultimate state is obtained.  

Computational studies of turbulent \RB convection have also explored scaling laws for vertical heat transport~\citep{verzicco2003numerical, amati2005turbulent, stringano2006mean, verzicco2008comparison}.  ~\citet{amati2005turbulent} performed direct numerical simulations of \RB convection within the \OB approximation up to $Ra = 10^{14}$ at $Pr=0.7$.  They found a scaling of $Nu\sim Ra^{1/3}$.  Numerical studies also seek to explain large scale flow structures~\citep{stringano2006mean} as well as discrepancies between experimental and numerical results~\citep{verzicco2008comparison}.  The scaling laws from the computational studies just mentioned are derived from fully turbulent flow fields which permits contact with the original work by~\citet{malkus1954heat} as well as with the GL theory.  Several data sets from numerical and experimental studies can be found in~\citet{stevens2013unifying} along with comparison to the GL theory.  Other computational studies have explored the effect of $Pr$ on heat transport. ~\citet{verzicco1999prandtl} and~\citet{breuer2004effect} performed simulations of \RB convection in order to discern $Pr$ effects on the flow field.  ~\citet{verzicco1999prandtl} performed simulations in a cylindrical cell with diameter to height ratio $D/H \equiv \Gamma = 1$ whereas~\citet{breuer2004effect} performed simulations in a box with $\Gamma = 2$.  Both studies found a threshold in $Pr$ at $Pr \approx 0.3$ above which $Nu$ displays very little variation with $Pr$.  Below that threshold the flow is dominated by a large scale wind and $Nu$ is highly dependent upon $Pr$.

In this work, we numerically compute steady solutions to the \OB equations that optimize vertical heat transport.  Thus we make contact with the upper bound theories mentioned above. However, our solutions exactly satisfy the \OB equations and do not include any approximations other than those introduced for numerical discretization.  Previous work has explored such solutions at $Pr=7$~\citep{waleffe2015heat} as well as for porous medium convection~\citep{wen2015structure}.  Here we consider the two-dimensional problem with no-slip boundary conditions for $Pr = 1, 4, 7, 10, 100$ and $Ra \lesssim 10^9$.  We discuss the structures that contribute to optimal vertical heat transport.  In particular, we explore the effect of $Pr$ on the optimal solutions and its impact on scaling laws.  A large body of early theoretical work considered \RB convection at $Pr =7$ in a domain of length $2\pi/\alpha$ where $\alpha = 1.5585$ is the wavenumber to which the conduction state is unstable for no slip boundary conditions~\citep{drazin2004hydrodynamic}.  We refer to solutions with that wavenumber $\alpha = 1.5585$ as ``primary'' solutions; these are the first nonlinear steady solutions that bifurcate from the conduction state.  However the optimum heat transport solution corresponds to a wavenumber that generally increases with $Ra$, starting from the fundamental wavenumber $\alpha = 1.5585$ at $Ra=1708$.  For some Prandtl numbers, we find the possible emergence of a 2nd local optimum transport solution and possible abrupt transitions in the wavenumber of the global optimum steady solution, from one local optimum to another.  In spite of these significant jumps in optimum wavenumber for some $Pr$, we do not find any significant $Pr$ dependence in the $Nu\lr{Ra}$ scaling.  However, a significant $Pr$ dependence is observed in determining precisely which structures lead to the optimum in $Nu$; in particular, $Pr$ controls the length scales of these structures.  Interestingly, $Pr \approx 7$ (water) represents a demarcation between fluids that require smaller wavenumbers and those that require larger wavenumbers to optimize vertical heat transport.

The remainder of the paper is as follows.  In section~\ref{sec:background} we introduce the governing equations, boundary conditions, and dimensionless parameters.  Section~\ref{sec:computation} discusses the computational method used to find fixed points and optimal solutions.  The main results are presented in section~\ref{sec:results}, followed by concluding remarks in section~\ref{sec:conclusions}.

\section{Background}\label{sec:background}

\subsection{Governing Equations}
We consider a fluid under the influence of gravity and confined between two infinite horizontal planes situated at $y = \pm h$ so that the channel height is $H=2h$.  The \OB approximation is employed wherein compressibility effects (i.e. density variations) are only significant in the buoyancy term.  The evolution of the velocity field $\displaystyle \bld{u}\lr{\bld{x}, t} = \lr{u, \ v}$ in a rectangular domain with cartesian coordinates $\displaystyle \bld{x} = \lr{x, \ y}$ is given by
\begin{align}
  \pdeone{\bld{u}}{t} + \bld{u}\cdot\nabla\bld{u} + \nabla P &= \nu\nabla^{2}\bld{u} + \alpha\ss{V}gT\widehat{\bld{y}} \label{eq:imom}\\
  \div\bld{u} &= 0 \label{eq:incomp}
\end{align}
where $P\lr{\bld{x}, t}$ is the kinematic pressure, $\nu$ is the kinematic viscosity, $\alpha\ss{V}$ is the volumetric thermal expansion coefficient, $g$ is the acceleration of gravity, and $\widehat{\bld{y}}$ is the unit vector in the vertical direction.  The pressure and the horizontal velocity can be eliminated by taking the double curl of equation~\eqref{eq:imom}.  A projection in the vertical direction then yields the vertical velocity equation
\begin{align}
  \pdeone{\nabla^{2}v}{t} + \pdeone{}{x}\lr{u\nabla^{2}v - v\nabla^{2}u} = \nu\nabla^{2}\nabla^{2}v + \alpha\ss{V}g\pden{T}{x}{2}. \label{eq:vmomp}
\end{align}
The horizontal velocity is recovered from the continuity equation \eqref{eq:incomp}.  The temperature, $\displaystyle T\lr{\bld{x},t}$, is obtained from the thermal energy equation that reduces to an advection-diffusion equation for the temperature, 
\begin{align}
  \pdeone{T}{t} + \bld{u}\cdot \nabla T &= \kappa\nabla^{2}T
\label{eq:energy}.
\end{align}
In equation~\eqref{eq:energy}, $\kappa$ is the thermal diffusivity.  Note that in general equations~\eqref{eq:vmomp}-~\eqref{eq:energy} must be supplemented by an equation for the horizontal average of the horizontal component of velocity which we refer to as the mean flow $\bar{u}\lr{y,t}$.  It is expected that the mean flow will decrease vertical heat transport and therefore optimal solutions for vertical heat transport will result when the mean flow is zero~\citep{kerswell2001new}.  We therefore directly set the mean flow to zero and impose mirror symmetry about the vertical plane $x=0$ so that the solutions maintain zero-mean.  Mirror symmetry is expressed mathematically as $[u,v,T]\lr{x,y,t} = [-u,v,T]\lr{-x,y,t}$.

In order to fully determine the velocity and temperature fields, equations~\eqref{eq:vmomp} and \eqref{eq:energy} must be augmented with boundary conditions.  We impose periodic boundary conditions in the horizontal direction for both fields.  At the top and bottom surfaces we prescribe uniform cold and hot temperatures, respectively,
\begin{align}
  T\lr{x,\pm h, t} = \mp T_{s}, \qquad T_{s} > 0.
\end{align}
No-slip conditions on the velocity field, $u=0$, $v=0$, are imposed at each surface so that
\begin{align}
  v\lr{x,\pm h, t} = 0, \quad \pdeone{v}{y}\lr{x,\pm h, t} = 0, \label{eq:noslip}
\end{align}
where the second equation follows from continuity.

The primary diagnostic quantity is the net heat flux through the top and bottom plates
\begin{equation}
  \mathcal{H}_{\pm}(t) \doteq -\kappa\left.\odeone{\overline{T}}{y}\right|_{y=\pm h} 
  \end{equation}
  where the overbar $\overline{(\cdot)}$ denotes a horizontal average. In statistically steady state,  $\mathcal{H}_{+}=\mathcal{H}_{-}=\mathcal{H}$ is constant. The Nusselt number $Nu$ is the ratio of the net heat flux $\mathcal{H}$ to the conductive heat flux $\mathcal{H}_0=\kappa \, T_s/h$, thus
 \begin{equation}
  Nu = \frac{\mathcal{H}}{\kappa \, T_s/h} = 
  -\frac{h}{T_s} \left.\odeone{\overline{T}}{y}\right|_{y=\pm h}.
  \label{eq:Nudim}
  \end{equation}

\subsection{Nondimensionalization}
In equations~\eqref{eq:vmomp}-\eqref{eq:energy} we nondimensionalize lengthscales by the 
half-height $h$.  The temperature is nondimensionalized by the half-temperature difference between the bottom and top plates, $\Delta T/2=T_{s}$.  The time scale is chosen as $\tau = \sqrt{h/g'}$ where $g' = g\alpha\ss{V}T_{s}$ is the reduced gravity and $2\tau$ is the time that it takes a blob of fluid at temperature $-T_{s}$ to free fall from the top wall to the bottom wall.  A velocity scale is then $U = h/\tau$ which is the average free fall speed.  The resulting nondimensional equations are
\begin{align}
  \pdeone{\nabla^{2}v}{t} + \pdeone{}{x}\lr{u\nabla^{2}v - v\nabla^{2}u} &= \nund\nabla^{2}\nabla^{2}v + \pden{T}{x}{2} \label{eq:vmom-nondim} \\
  \div\bld{u} &= 0 \label{eq:cont-nondim} \\
  \pdeone{T}{t} + \bld{u}\cdot\nabla T &= \kand\nabla^{2}T \label{eq:energy-nondim}
\end{align}
with boundary conditions
\begin{align}
              &T\lr{\pm 1} = \mp 1 \\
  v\lr{x,\pm 1,t} &= 0, \quad \pdeone{v}{y}\lr{x,\pm 1,t} = 0. \label{eq:noslip-nondim}
\end{align}
This nondimensionalization leads to dimensionless diffusivities
\begin{align}
  \nund    = \frac{\nu}{\sqrt{g'h^{3}}} \qquad \text{and} \qquad \kand = \frac{\kappa}{\sqrt{g'h^{3}}}
\end{align}
which are related to the Prandtl number, $Pr = \nu/\kappa$, and the Rayleigh number, $Ra = 2g'H^{3}/\nu\kappa$, through
\begin{align*}
  \nund = \sqrt{\frac{16Pr}{Ra}} \qquad \text{and} \qquad \kand = \sqrt{\frac{16}{RaPr}}.
\end{align*}

\subsection{Heat Transport}
In our non-dimensionalization the Nusselt number, \eqref{eq:Nudim} is simply
\begin{equation}
Nu= -\left.\odeone{\overline{T}}{y}\right|_{y=-1}.
\end{equation}
Integrating \eqref{eq:energy-nondim} over horizontal planes, using continuity to write $\bld{u}\cdot\nabla T = \nabla \cdot \left(\bld{u} T\right)$,  yields  
\begin{align}
  \odeone{}{y}\lr{-\kand\odeone{\overline{T}}{y}+  \overline{vT}} = 0
  \label{eq:energy-hor-ave}
\end{align}
assuming a statistical steady state (steady horizontal averages). This indicates that the net heat flux through horizontal planes is constant  
\begin{align}
  -\kand\odeone{\overline{T}}{y}+  \overline{vT} =
  -\kand \left.\odeone{\overline{T}}{y}\right|_{y=-1} = \kand \; Nu
  \label{eq:heat-flux}
\end{align}
since $v=0$ at the bottom plate.  Averaging \eqref{eq:heat-flux} over the channel height, from $\overline{T}=1$ at $y=-1$ to $\overline{T}=-1$ at $y=+1$, then gives
\begin{align}
  Nu = 1 + \frac{\left<vT\right>}{\kand} \label{eq:Nu}
\end{align}
where $\left<\cdot\right>$ represents a volume average.  In the pure conduction state $v = 0$ and $Nu = 1$.  It is well-known, however, that the conduction state is unstable above a critical $Ra$ of $1708$~\citep{jeffreys1928some, pellew1940maintained, reid1958some}.  Above this $Ra$ the solution bifurcates from the conduction state to a steady convection state. It can be shown from the kinetic energy equation that $\langle vT \rangle > 0$ whenever $\bld{u} \ne 0$ (again in statistically steady state).  The onset of convection thus acts to enhance heat transport and $Nu > 1$ for $Ra > 1708$.  In this work, we search for steady solutions that optimize $Nu$ as a function of $\alpha$ at a given $Ra$.  We find that $Nu_{opt} \sim Ra^{\gamma}$ for sufficiently large $Ra$ at fixed $Pr$.

\section{Computational Method}~\label{sec:computation}
We seek steady, possibly unstable solutions of equations~\eqref{eq:vmom-nondim}-\eqref{eq:energy-nondim} that maximize $Nu$ \eqref{eq:Nu} over $\alpha$. To accomplish this task, we have implemented a continuation method to find fixed points that uses $Ra$ as the continuation parameter.  Our implementation is similar to that used by~\citet{sanchez2004newton} and \citet{viswanath2007recurrent} to track periodic orbits in annular thermal convection and plane Couette turbulence.~\citet{net2003stationary} used a similar algorithm to find stationary states in annular \RB convection.

At each $Ra$ we find steady solutions of various horizontal wavelengths, and then determine which wavelength results in the maximum vertical heat transport.  We call this solution the optimal solution although it is only guaranteed to be a locally optimal steady solution.  The richness of the \OB equations means that many types of solutions exist.  For example, as will be shown in section~\ref{sec:CS}, our optimal solutions have a single thermal plume emanating from the lower surface.  However, it is feasible for other solutions to exist such as two opposing plumes with one plume emanating from the bottom surface and the other from the top surface.  We do not expect such opposing plumes to result in optimal vertical heat transport but other non-steady or 3D flows might possibly yield higher transport.

The length of the channel in our simulations is given by $L=2\pi/\alpha$ where the wavenumber $\alpha$ characterizes the largest permissible wavelength in the channel.  We use the \flowmap algorithm outlined in section~\ref{sec:flowmap} to find fixed points.  Before discussing the fixed point algorithm, we briefly review the method we used for solving the fourth-order system (equation~\eqref{eq:vmom-nondim}).

\subsection{Solution of the \OB Equations}\label{sec:OBeqs-sol}
The momentum equation for the vertical velocity field (equation~\eqref{eq:vmom-nondim}) is fourth order and therefore requires the four boundary conditions stated in equation~\eqref{eq:noslip-nondim}.  In order to efficiently solve these equations we follow a procedure in which an auxiliary variable, $\phi$, is introduced that reduces the order of the PDE from fourth-order to second-order~\citep{kim1987turbulence}.  With $\phi = \nabla^{2}v$ the vertical velocity can be determined from,
\begin{align}
  \pdeone{\phi}{t} + \pdeone{}{x}\lr{u\phi - v\nabla^{2}u} &= \nund\nabla^{2}\phi + \pden{T}{x}{2} \label{eq:phi-eq} \\
  \nabla^{2}v &= \phi \label{eq:vphi-eq}.
\end{align}
Although we are determining steady states, our algorithm still requires an accurate time-integrator (see section~\ref{sec:flowmap}).  In this work we use an adaptive implicit-explicit (IMEX) Runge-Kutta (RK) method that is third order accurate.  In particular, we use a $\lr{3,4,3}$ (3 implicit stages, 4 explicit stages, third order accurate) IMEX-RK scheme~\citep{ascher1995implicit}.  Details on satisfaction of the boundary conditions and the time integration implementation for our particular problem are provided in appendix~\ref{app:BCTI}.  The spatial discretization is accomplished using a Fourier basis in the horizontal, periodic direction and a non-uniform finite difference scheme in the vertical direction based on Lagrange polynomials.  The $2/3$ dealiasing rule is applied in the horizontal direction to remove aliasing errors associated with the nonlinear terms.  We checked the numerical resolution of our simulations in several ways.  For low $Ra$ we were able to perform a mesh-refinement study in which we doubled the spatial resolution.  We then ensured that the $Nu$ between the fine and coarse solutions agreed.  In general, our simulations had at least $10-20$ points in the boundary layers which satisfies the requirements suggested by~\citet{grotzbach1983spatial}.  A third resolution requirement that was sometimes employed was to check the horizontal energy spectra of the velocity and temperature fields to make sure that energy did not pile up in any wavenumbers. Finally, our results were checked against the $Pr=7$ results of \citet{waleffe2015heat} and some of the results presented here for $Pr=1$ and $Pr=7$ were obtained with code 2 in that reference.

\subsection{Flow Map Algorithm and Fixed Points}\label{sec:flowmap}
We now describe the \flowmap algorithm used to determine fixed points of this system.  Let $X = \lr{\phi, T}$ be a vector of solutions.  We introduce a spatial discretization of the equations of interest (i.e. equations~\eqref{eq:phi-eq} and~\eqref{eq:energy-nondim}) as described in section~\ref{sec:OBeqs-sol} from which we obtain a time-dependent system of ordinary differential equations,
\begin{align}
  \odeone{X}{t} = F\lr{X}. \label{eq:Fx}
\end{align}
One way to find steady states is to directly determine fixed points $X$ such that 
\begin{align}
  F\lr{X} = 0. \label{eq:Fxeqzero}
\end{align}
Such methods represent direct-to-steady-state methods.  An alternative way of determining fixed points is to use the flow map.  Let $X$ be any initial state and $X_{T}$ the corresponding solution at time $T$.  If $X$ is a fixed point then the difference
\begin{align}
  G_{T}\lr{X} = \frac{X_{T} - X}{T} \label{eq:flowmap}
\end{align}
should be zero.  Hence the goal is to find $X$ such that
\begin{align}
  G_{T}\lr{X} = 0. \label{eq:GTeqzero}
\end{align}
For a small integration time $T$ the difference $X_{T} - X$ will be small.  We divide by $T$ so that $G_{T}(X) \to F(X)$ as $T\to 0$.  On the other hand, it is expected that $G_{T}\lr{X} \to 0$ as $T\to \infty$, assuming that $|X_T-X|$ is bounded.  Choosing $T$ is a matter of experience for a particular problem, but it should be relatively small compared to the time scales of the modes that prevent rapid relaxation to steady state. 

We have studied the effects of the \flowmap approach on solving a large linear system of equations $\odeone{X}{t} = \mathbf{A}X - \mathbf{b}$, where the eigenvalues of $\mathbf{A}$ mimic those of problems of interest, such as diffusion.  The \flowmap acts as a preconditioner for the resulting Jacobian matrix: it can improve the condition number of the Jacobian matrix and redistribute the eigenvalues such that they cluster around the origin.  These observations are consistent with previous studies~\citep{sanchez2004newton} and indicate improvements for solving some linear systems of equations.  In particular, the generalized minimum residual (GMRES) method is known to converge well when the eigenvalues of the matrix are clustered around the origin~\citep{trefethen1997numerical}.

\subsection{Details of the Flow Map Algorithm}\label{sec:alg-details}
Equation~\eqref{eq:GTeqzero} represents a large system of equations with approximately $N$ unknowns where $N$ increases with $Ra$.  In this study $N \approx 10^6$ for the largest values of $Ra$ used.  Equation~\eqref{eq:GTeqzero} is typically solved using some variant of Newton's method.  The full version of Newton's method can be difficult to implement and expensive to use.  Difficulties include forming the full Jacobian matrix and efficiently solving the resulting matrix system.  Sophisticated approaches exist to help make Newton's method more cost-effective.  These include preconditioning the Jacobian matrix, inexact Newton methods based on Krylov subspaces, and Jacobian-free methods~\citep{knoll2004jacobian}.

In the present work, we solve~\eqref{eq:GTeqzero} by means of a Jacobian-free Newton-Krylov method.  Newton's method results in the linear system,
\begin{align}
 \left.\odeone{G_{T}}{X}\right|_{X=X_{k}}\Delta X = -\left.G_{T}\right|_{X=X_{k}} \label{eq:linsys}
\end{align}
where $X_{k}$ represents the solutions at nonlinear iteration $k$ and $\displaystyle \Delta X = X_{k+1}-X_{k}$.  Rather than forming the full Jacobian, a forward finite difference scheme is used to approximate the action of the Jacobian on $\Delta X$~\citep{knoll2004jacobian}.  The GMRES method is then used to solve the resulting linear system~\citep{saad1986gmres, trefethen1997numerical}.  We note that it is not necessary to form the full Jacobian (or even its approximation) because the GMRES algorithm only requires the action of a matrix on the solution.  Each iteration of the GMRES method solves equation~\eqref{eq:linsys} with
\begin{align}
  \left.\odeone{G_{T}}{X}\right|_{X=X_{k}}\Delta X \approx \frac{G_{T}\lr{X_{k} + \epsilon\Delta X} - G_{T}\lr{X_{k}}}{\epsilon} \label{eq:jacapprox}
\end{align}
with the choice
\begin{align}
  \epsilon = \frac{\sqrt{\epsilon_{\text{machine}}}}{\|\Delta X\|^{2}}\text{max}\lr{X_{k}\cdot\Delta X, \|\Delta X\|} \label{eq:epsi}
\end{align}
where $\epsilon_{\text{machine}}$ is the machine tolerance and $\|\cdot\|$ is the $L_{2}$ norm.  Several possibilities exist for the choice of $\epsilon$~\citep[see][pg. 363]{knoll2004jacobian} but the main point is that it must be small enough to get an accurate derivative in~\eqref{eq:jacapprox} but not so small that floating point errors pollute the solution~\citep[section 5.4]{dennis1996numerical}.

We conclude this section with a few comments and observations based on our experience with the \flowmap algorithm.  The fact that the \flowmap algorithm redistributes the eigenvalues of the Jacobian matrix also has benefits when using the GMRES method to solve the linear system~\eqref{eq:linsys}. To understand the latter, recall that the basis vectors for the Krylov subspace are formed from a power method, and that the power method focuses on the largest eigenvalue of the system.  Since the \flowmap algorithm results in a larger separation between the largest eigenvalue and the other eigenvalues, the power method converges faster.  Hence, a smaller Krylov subspace is required for the linear solves.

In terms of computation time, there appears to be an optimal integration time $T$ that minimizes the wall-clock time of the simulation.  This optimal time is dependent upon the problem at hand.  For example, increasing $T$ causes all time-integrations to take longer.  On the other hand, increasing $T$ allows for better preconditioning and results in a much reduced time for solving the linear system.  For the current problem, we have observed that $T \approx 0.6$ results in the quickest computation time for low $Ra$ around $10^{5}$.  We have used $T=2.0$ throughout for both high and low $Ra$. Some of the results presented herein for $Pr=1$ and $Pr=7$ for $Ra \geq 10^6$ were computed with code 2 in \citet{waleffe2015heat} which is based on spectral Chebyshev integration and does not use GMRES.

\section{Results}\label{sec:results}
In this section, we present steady solutions to the \OB equations that maximize heat transport.  We consider Prandtl numbers $Pr$ ranging from $1$ to $100$ and note that these values roughly correspond to real fluids.  For example, $Pr=7$ is often used for water and engine oil has $Pr$ of roughly $100$.  In subsection~\ref{sec:MM}, we explore the effects of both $Pr$ and the horizontal wavenumber $\alpha$ on optimal heat transport.  Subsection~\ref{sec:SL} investigates the impact of $Pr$ on scaling laws, while subsection~\ref{sec:CS} discusses the coherent structures that give rise to the optimal solutions.  The mean temperature profiles associated with the optimal solutions are considered in subsection~\ref{sec:MTP}.

\subsection{Multiple Local Maxima and $Pr$ Effects}~\label{sec:MM}
\begin{figure}
  \centerline{\includegraphics[width=0.85\textwidth]{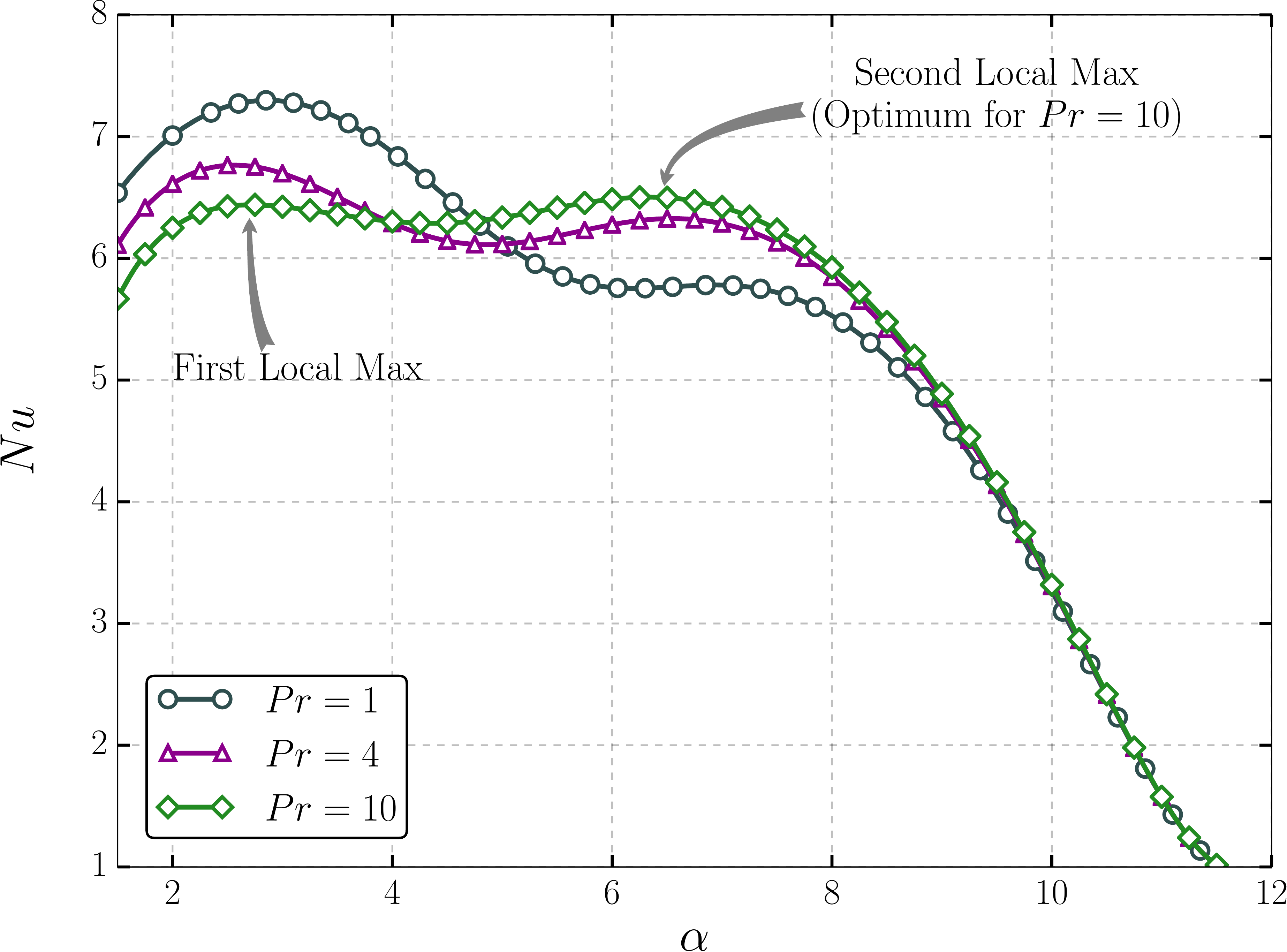}}
  \caption{Variation of $Nu$ with horizontal wavenumber $\alpha$ at $Ra = 3\times 10^{5}$ for $Pr = 1, 4, 10$.}
  \label{fig:Nu-alpha}
\end{figure}

The variation of heat transport $Nu$ with wavenumber $\alpha$ is shown in figure~\ref{fig:Nu-alpha} for $Pr = 1$, $Pr = 4$ and $Pr = 10$ at fixed $Ra = 3 \times 10^5$.  We note that two local maxima emerge for these three fluids.  Thus, there are two scales that locally maximize heat transport.  The solution corresponding to the global maximum, hereafter called the optimal solution, occurs at different length scales depending upon the fluid under consideration.  For example, the global maximum occurs at a small wavenumber $\alpha$ for $Pr=1$, whereas it occurs at a larger $\alpha$ for $Pr = 10$.  We have indicated in the figure the `first' and `second' local maxima for $Pr=10$.  When we refer to the first local maximum, we have in mind the maximum that occurs at smaller wavenumber $\alpha$, whereas when we refer to the second local maximum, we mean the maximum that occurs at larger wavenumber $\alpha$.

\begin{figure}
  \centering
  \begin{subfigure}[t]{0.475\textwidth}
    \includegraphics[width=\textwidth]{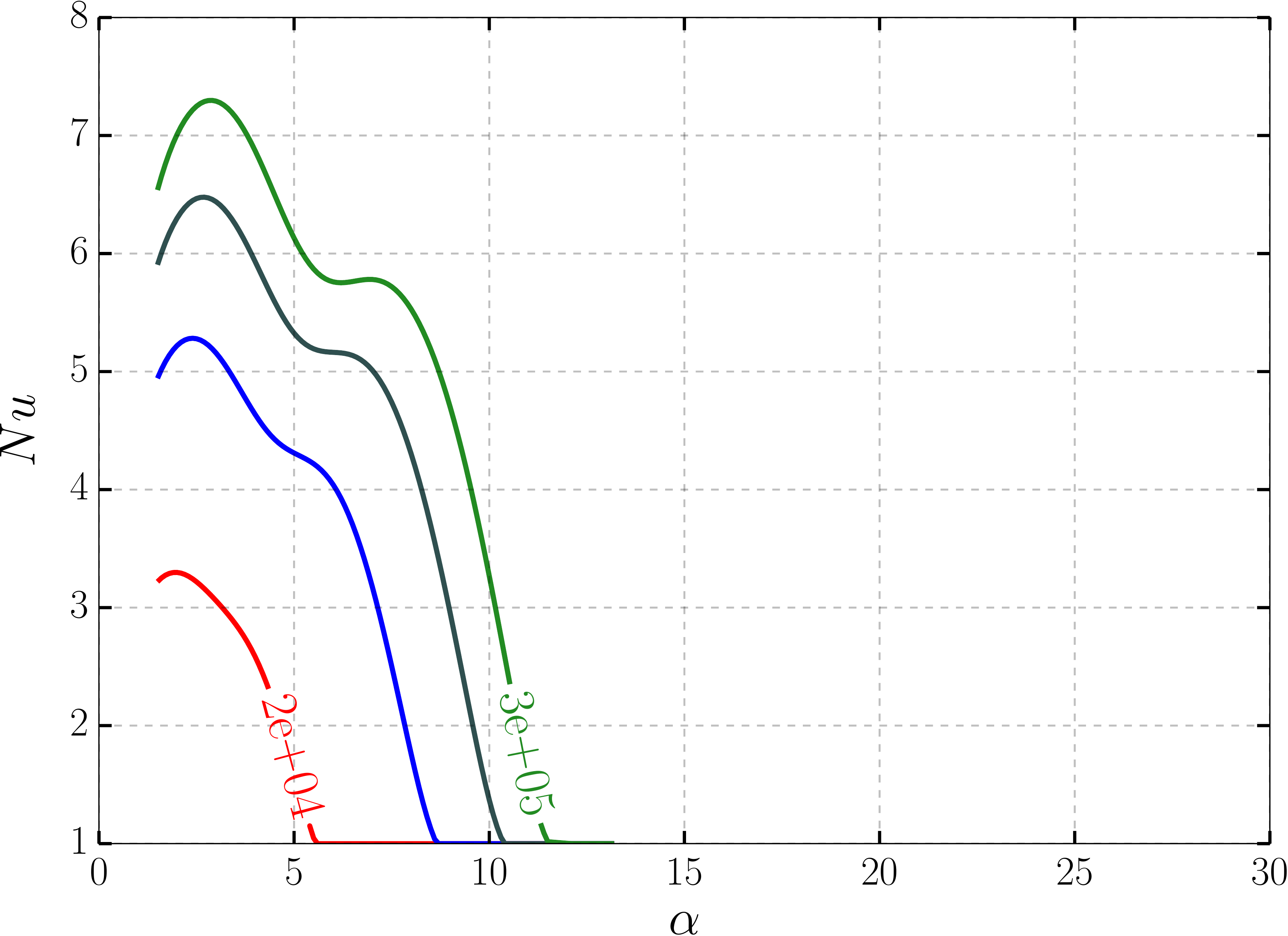}
    \begin{tikzpicture}[overlay]
      \node [scale=1.25] at (5.75,1.2) {(a)};
    \end{tikzpicture}
    \phantomcaption\label{fig:Nu-contours-Pr1}
  \end{subfigure}
  ~
  \begin{subfigure}[t]{0.475\textwidth}
    \includegraphics[width=\textwidth]{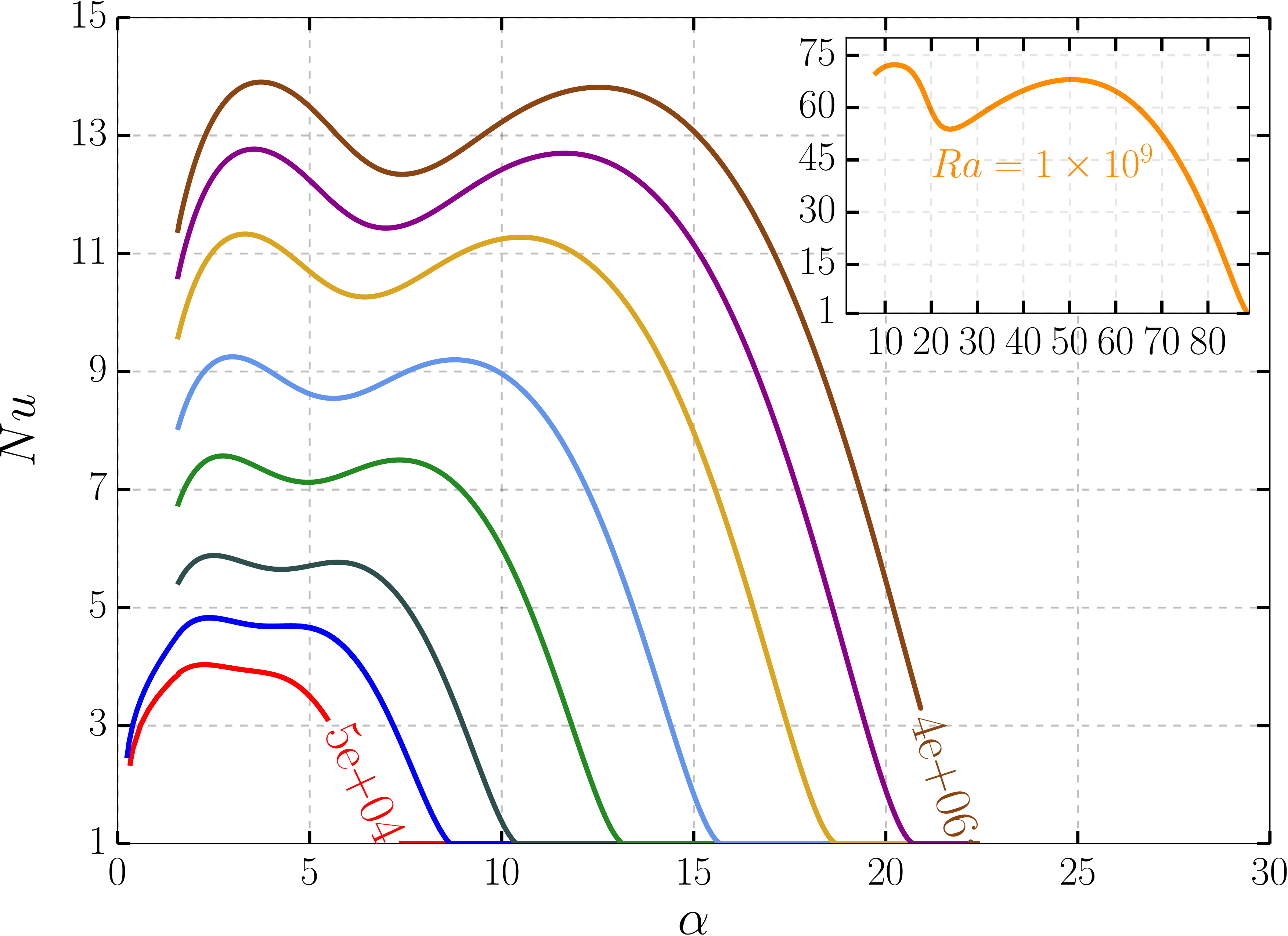}
    \begin{tikzpicture}[overlay]
      \node [scale=1.25] at (5.75,1.2) {(b)};
    \end{tikzpicture}
    \phantomcaption\label{fig:Nu-contours-Pr7}
  \end{subfigure}
  \\[1.0em]
  \begin{subfigure}[t]{0.475\textwidth}
    \includegraphics[width=\textwidth]{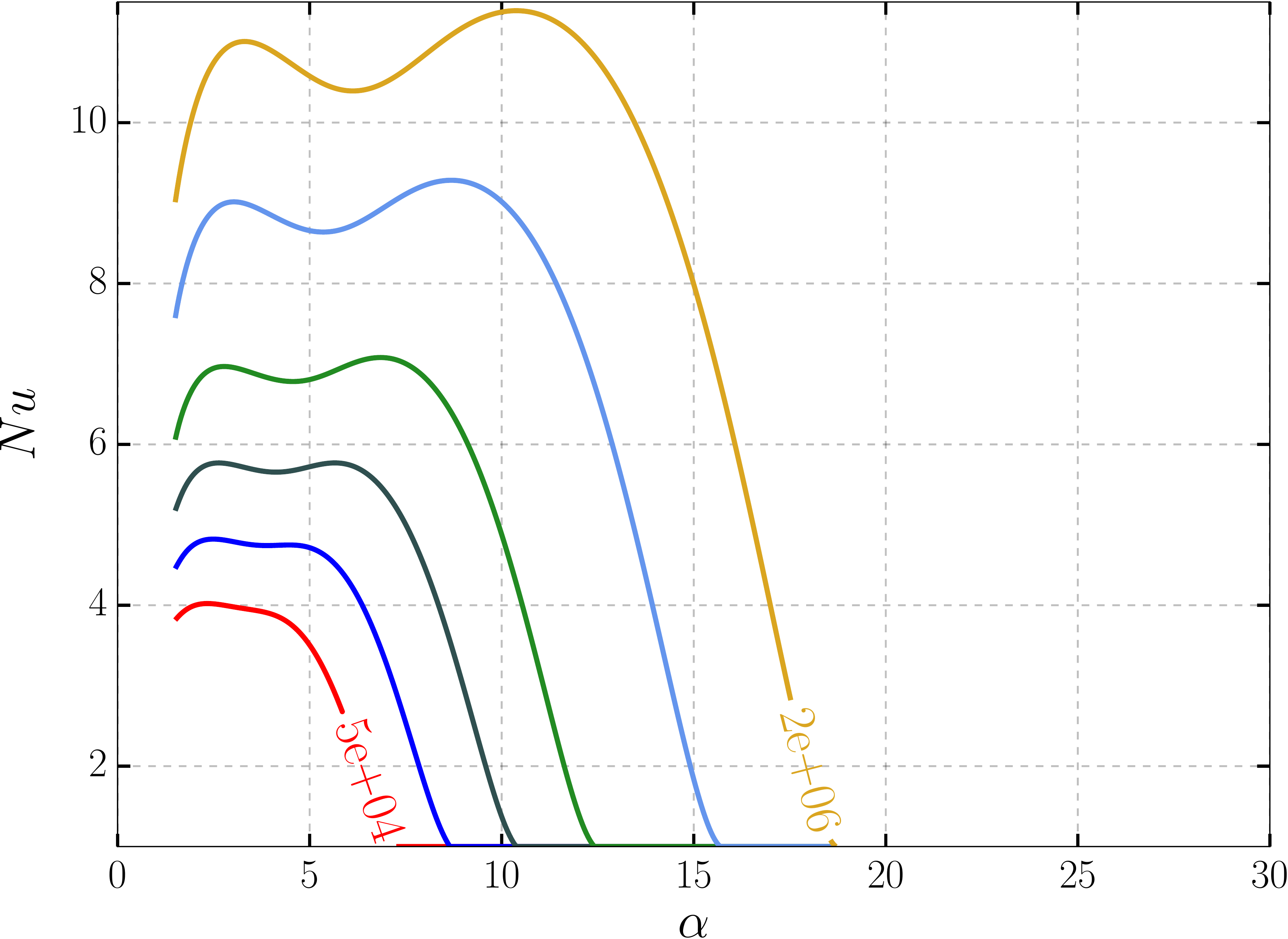}
    \begin{tikzpicture}[overlay]
      \node [scale=1.25] at (5.75,1.2) {(c)};
    \end{tikzpicture}
    \phantomcaption\label{fig:Nu-contours-Pr10}
  \end{subfigure}
  ~
  \begin{subfigure}[t]{0.475\textwidth}
    \includegraphics[width=\textwidth]{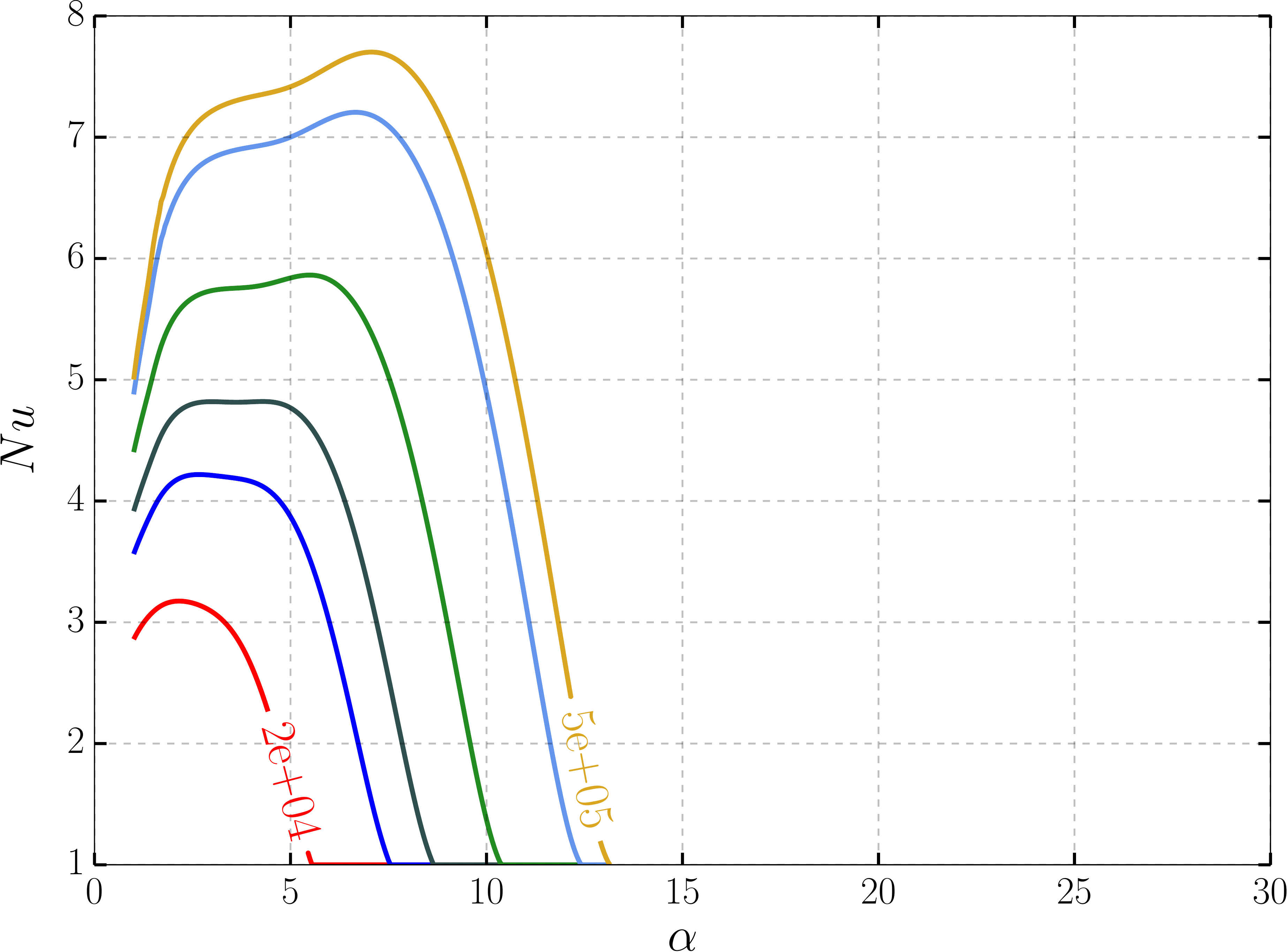}
    \begin{tikzpicture}[overlay]
      \node [scale=1.25] at (5.75,1.2) {(d)};
    \end{tikzpicture}
    \phantomcaption\label{fig:Nu-contours-Pr100}
  \end{subfigure}
  \caption{Variation of $Nu$ with horizontal wavenumber $\alpha$ for various $Ra$ at 
  (\subref{fig:Nu-contours-Pr1}) $Pr=1$,  
  (\subref{fig:Nu-contours-Pr7}) $Pr=7$, 
  (\subref{fig:Nu-contours-Pr10}) $Pr=10$ and 
  (\subref{fig:Nu-contours-Pr100}) $Pr=100$. 
The first and last $Ra$ contour values are marked in each plot.  Note that the color scheme between different plots is different as is the $Nu$ scale, however the $\alpha$ scale is identical.}
  \label{fig:Nu-contours}
\end{figure}
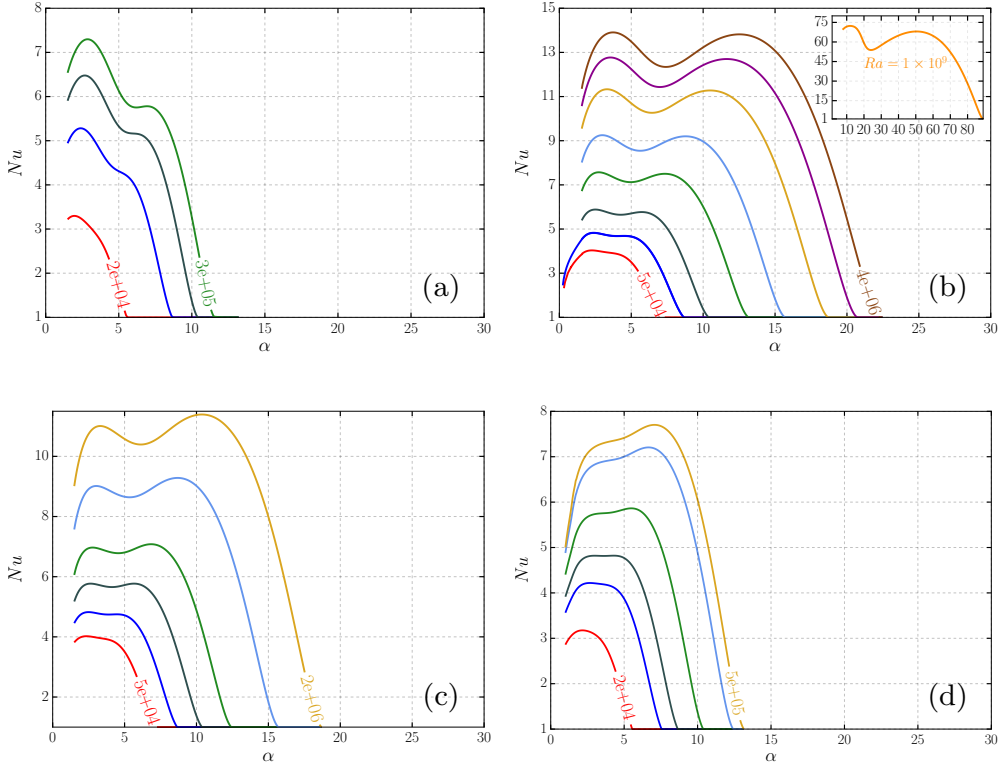

An intriguing behavior is illuminated upon inspection of other $Ra$ and $Pr$.  As indicated in figure~\ref{fig:Nu-contours}, it is not always the case that two local maxima are present.  Four different scenarios are described for $Pr=1$ (\ref{fig:Nu-contours-Pr1}), $Pr=7$ (\ref{fig:Nu-contours-Pr7}), $Pr=10$ (\ref{fig:Nu-contours-Pr10}) and $Pr=100$ (\ref{fig:Nu-contours-Pr100}).\\

\makebox[44pt][l]
{Figure~\ref{fig:Nu-contours-Pr1};}
\begin{minipage}[t]{300pt}
$Pr = 1$; $2\times 10^{4} \leq Ra \leq 3 \times 10^5$: 
For $Ra < 10^{5}$ only one maximum exists.  However, for $Ra\geq 2\times 10^{5}$ we observe two local maxima.  The first maximum is the optimal solution and there is no indication (at least for $Ra < 3 \times 10^5$) that the second local maximum will become the global maximum.\\
\end{minipage} 

\makebox[44pt][l]
{Figure~\ref{fig:Nu-contours-Pr7};}
\begin{minipage}[t]{300pt}
$Pr = 7$; $5\times 10^{4} \leq Ra \leq 4\times 10^{6}$:
For $Ra < 10^{5}$ there is only one local maximum.  However, for $Ra\geq 10^{5}$ two local maxima are present.  Moreover, the second local maximum is growing, and by $Ra = 3\times 10^{6}$, the two local maxima are nearly identical, although they occur at different scales.  In this $Ra$-range, it is not clear whether the first local maximum or the second local maximum will ultimately be the optimal solution, and the issue must be settled by computing maxima at higher $Ra$.  We have performed these computations, and the inset presents the $Nu-\alpha$ behavior at $Ra = 1\times 10^{9}$.  We observe that at this $Ra$ the second local maximum has receded compared to the first local maximum.\\
\end{minipage}

\makebox[44pt][l]
{Figure~\ref{fig:Nu-contours-Pr10};}
\begin{minipage}[t]{300pt}
$Pr = 10$; $5\times 10^{4} \leq Ra \leq 2\times 10^{6}$:
Yet another scenario is observed, wherein the new feature is that the second local maximum actually becomes the global maximum above a certain $Ra$.  The second local maximum appears at $Ra\approx 9\times 10^{4}$ and grows relative to the first local maximum.  At $Ra\approx 2\times 10^{5}$, the second local maximum is equal to the first local maximum.  Beyond this $Ra$, the second local maximum becomes the global maximum.\\
\end{minipage}

\makebox[44pt][l]
{Figure~\ref{fig:Nu-contours-Pr100};}
\begin{minipage}[t]{300pt}
$Pr = 100$; $2\times 10^{4} \leq Ra \leq 5\times 10^{5}$:
The large $Pr=100$ case introduces a fourth type of behavior with $Ra$, where the first local maximum vanishes for high enough $Ra$.  Once again, a single local maximum is present for $Ra$ below a certain value, this time $Ra = 8.5 \times 10^{4}$.  A second local maximum appears at $Ra = 8.5\times 10^{4}$ and begins to grow relative to the first local maximum.  By $Ra = 10^{5}$, the first and second local maxima are equal.  Interestingly, by $Ra = 2\times 10^{5}$ the first local maximum has vanished and the second local maximum is the global maximum.\\
\end{minipage}

\noindent
For ease of discussion below, we use the term `coexistence interval' to refer to the $Ra$-interval in which two local maxima are present, recognizing that the interval changes with $Pr$.

\begin{figure}
  \centering
  \includegraphics[width=0.85\textwidth]{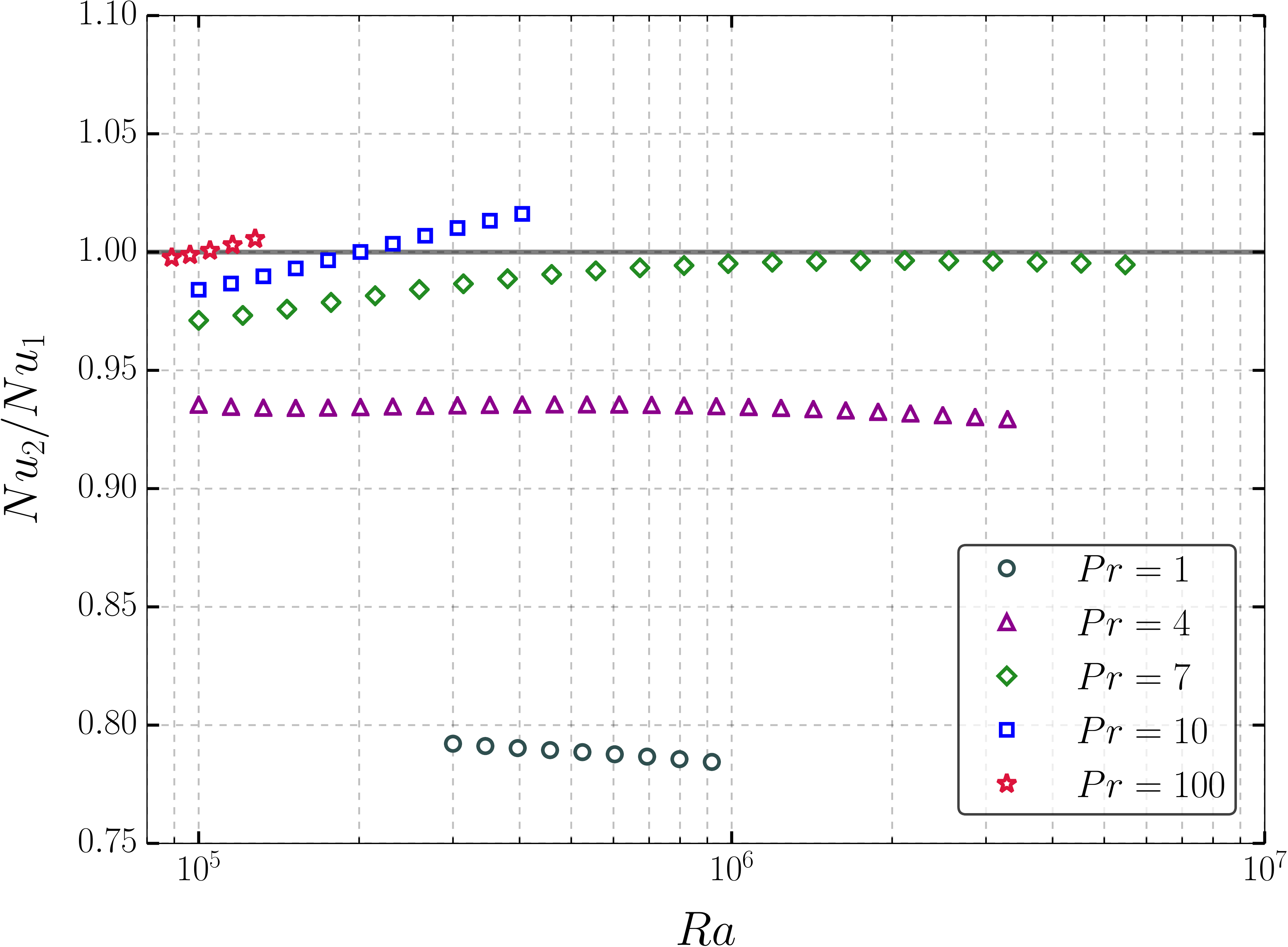}
  \caption{Ratio of local maxima as a function of $Ra$ for various $Pr$.}
  \label{fig:Nu-envelopes}
\end{figure}

Figure~\ref{fig:Nu-envelopes} summarizes the different scenarios we observed in the Prandtl number range $1 \leq Pr \leq 100$, for Rayleigh numbers $Ra < 10^7$. The figure shows the ratio of the second local maximum to the first local maximum $Nu\ss{2}/Nu\ss{1}$ as a function of $Ra$.  For $Pr=1$, $Pr=4$ and $Pr=7$ the second local maximum does not overtake the first maximum, whereas for $Pr=10$ and $Pr=100$ the second maximum eventually becomes the global maximum.  For $Pr = 7$, the two maxima are almost equal with the second local maximum becoming $99.6\%$ of the first local maximum at $Ra \approx 2 \times 10^6$ before slowing dropping off.  We also note that the coexistence region of the two maxima depends heavily on $Pr$.  This is most clearly observed for the $Pr=100$ case, in which the two maxima coexist only for a very small interval of $Ra$.  For $Pr=4$, $Pr=7$ and $Pr=10$ we did not observe termination of the coexistence region.  However, for $Pr=1$ and $Pr = 100$ the coexistence region was found to be finite, with the second (first) maximum disappearing for $Pr=1$ ($Pr = 100$).
Figure~\ref{fig:Nu-envelopes} implies the existence of two different parametric regions for optimal heat transport roughly partitioned by $Pr = 7$.  For $Pr>7$, the second local maximum overtakes the first local maximum to become the optimal solution and appears to remain the optimal solution.  Since the second local maximum occurs at larger $\alpha$, our calculations suggest that optimal solutions occur at relatively smaller scales for fluids with $Pr > 7$ at high-enough $Ra$.  Conversely, for $Pr\lesssim 7$, the second local maximum never overtakes the first local maximum, and therefore the first local maximum remains the optimal solution.  Thus, for fluids with $Pr\lesssim 7$, optimal solutions occur at the larger of the two length scales.  The $Pr$ dependence is significant in determining the flow structure that leads to optimal solutions.

\subsection{Scaling Laws and Impact of $Pr$}\label{sec:SL}
\begin{figure}
  \centering
  \includegraphics[width=0.85\textwidth]{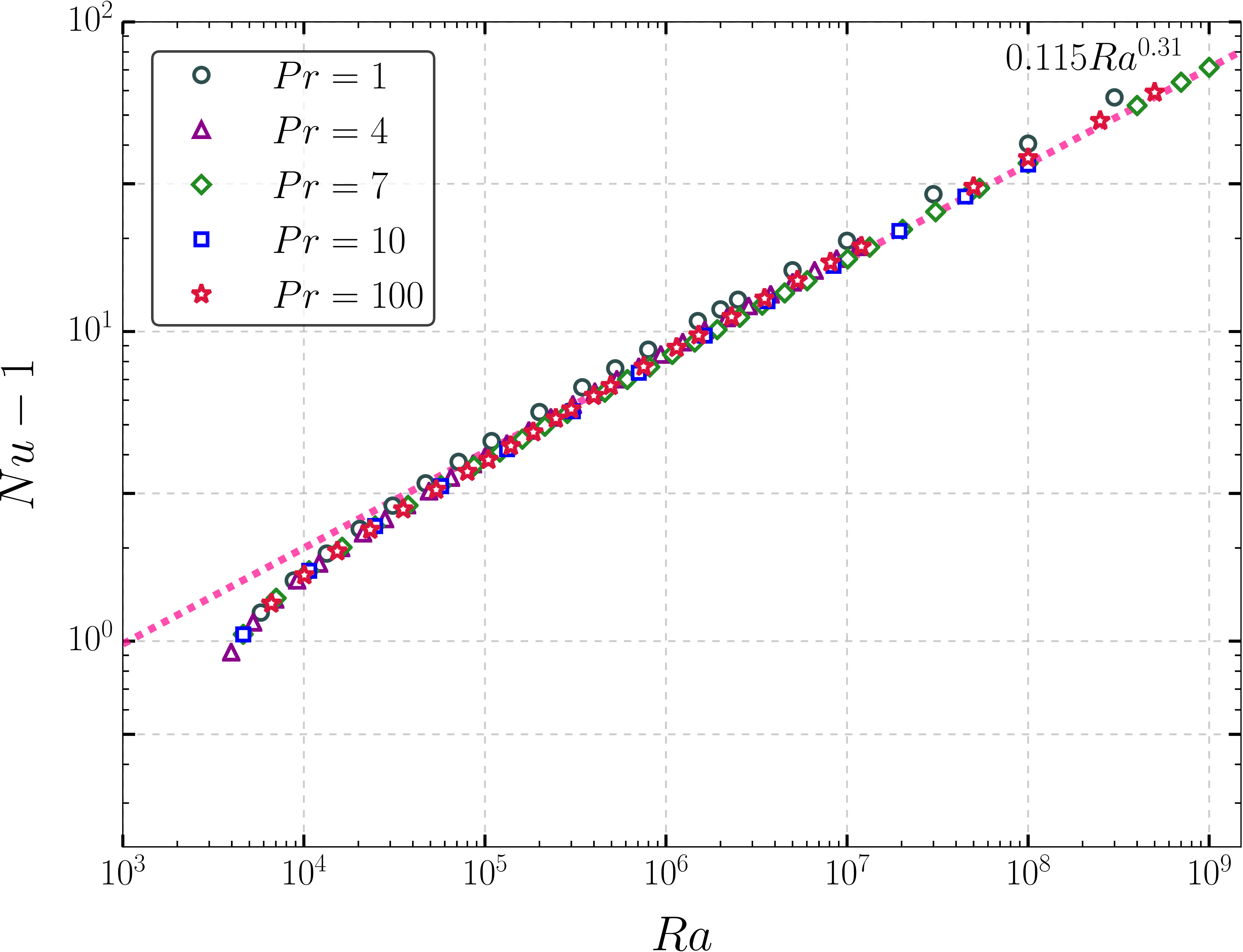}
  \caption{Optimal $Nu$ vs.\ $Ra$; global optimal data is shown and may correspond to distinct wavenumber.}
  \label{fig:Nu-Ra}
\end{figure}

\begin{figure}
  \centering
  \includegraphics[width=0.85\textwidth]{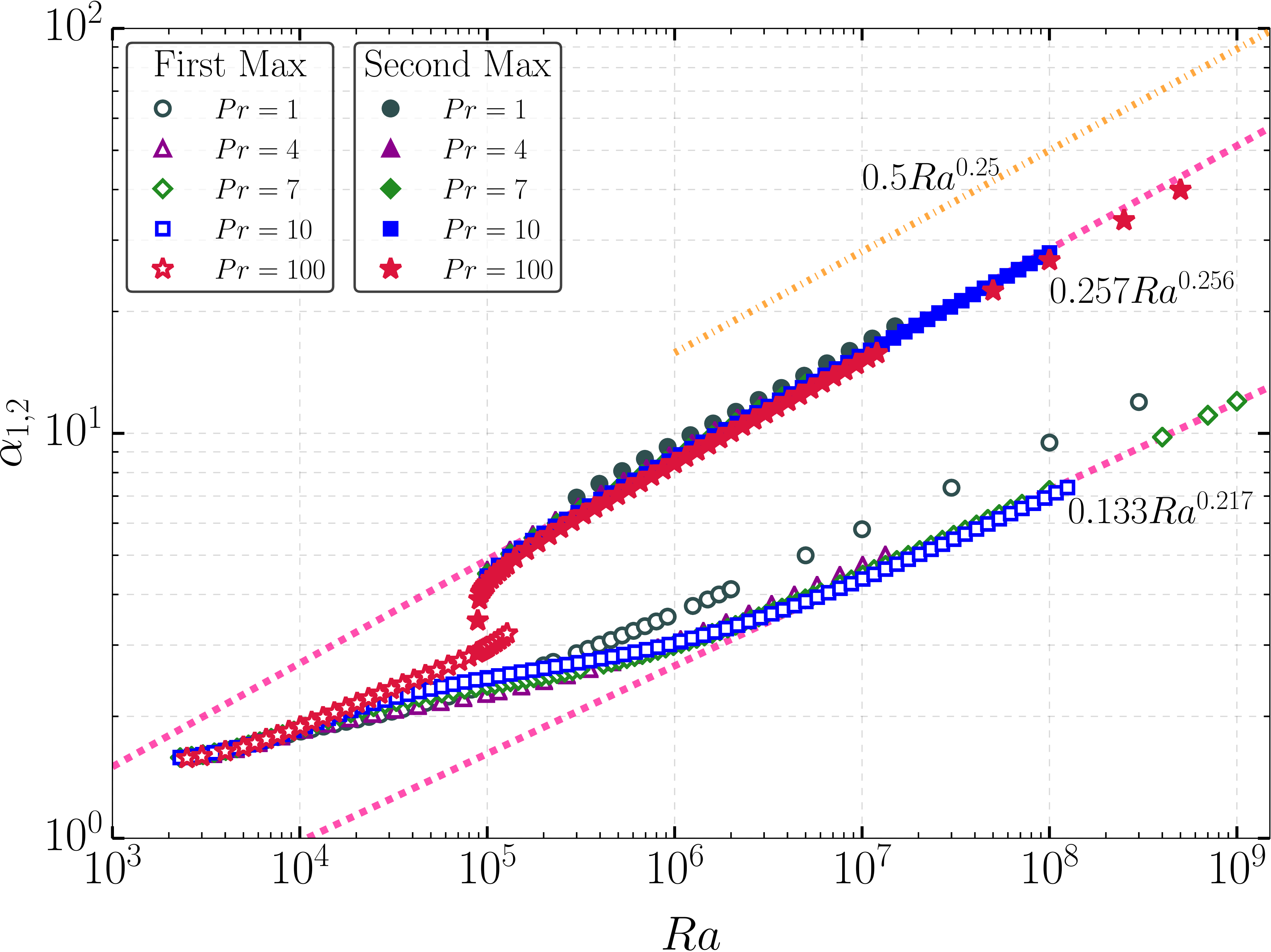}
  \caption{Maximal $\alpha$ vs.\ $Ra$; open markers indicate the first maximum and filled markers indicate the second maximum; the upper dashed-dot line represents the upper bound from linear stability theory.}
  \label{fig:alpha-Ra}
\end{figure}

Figures~\ref{fig:Nu-Ra} and~\ref{fig:alpha-Ra} present log-log plots of optimal $Nu$ and optimal $\alpha$ with $Ra$.  In figure~\ref{fig:Nu-Ra} we observe the dependence of the optimal $Nu$ on $Ra$ in the form $Nu\propto Ra^{\gamma}$ for $Ra > 10^5$.  The scaling law of $Nu = 0.115\, Ra^{0.31}$ was determined for $Pr = 7$ in~\citet{waleffe2015heat}, and is shown by the dashed line in figure~\ref{fig:Nu-Ra} for comparison.  This scaling was obtained from data in the range $10^{7}\leq Ra \leq 10^{9}$.  We observe that there is very little variation from this scaling with $Pr$ although the prefactor for $Pr=1$ is higher. 

Figure~\ref{fig:alpha-Ra} shows the dependence of the two maximal wavenumbers $\alpha\ss{1,2}$ on the Rayleigh number $Ra$. We begin our discussion by focusing on the open markers in figure~\ref{fig:alpha-Ra} (first maximum).  For $Pr = 100$, the first local maximum $\alpha_1$ vanishes after $Ra = 1.3\times 10^{5}$ and hence a clear scaling of $\alpha_1$ with $Ra$ cannot be determined.  However, for the other $Pr$ considered in this work, a scaling regime appears.  The scaling of $\alpha_1$ determined in~\citet{waleffe2015heat} for $Pr=7$ is shown in the dashed line: $\alpha_1 = 0.133 Ra^{0.217}$.  Once again, the scaling shows almost no change with $Pr$.  For $Pr = 1$ the power law of the scaling appears consistent with the other $Pr$ although it is likely that the prefactor is larger.  We note that for $Ra<10^{5}$, the $Ra$ dependence of $\alpha_1$ exhibits some wobbles, conjectured to be linked to the appearance/disappearance of coiling arms emanating from the central temperature plume of the optimal steady solutions~\citep{waleffe2015heat}.

Next we turn attention to the closed markers in figure~\ref{fig:alpha-Ra} corresponding to the second maximum with wavenumber $\alpha_2$.  A scaling regime emerges at $Ra\approx 10^{6}$,  and the scaling regime for these solutions is more complete because all fluids considered in this work exhibited a second local maximum in vertical heat transport that persisted for several decades of $Ra$. For $Pr=7$ we find $\alpha_2 \approx 0.257Ra^{0.256}$ with the current range of $Ra$ sampled by our data.  It can be shown from linear stability theory ($Nu=1$) that the upper wavenumber limit for marginal stability $\alpha_u$ scales as  $\alpha_u \sim 0.5Ra^{1/4}$ for large $Ra$.  This scaling is marked in the dash-dot line in figure~\ref{fig:alpha-Ra}. $Nu$ drops to 1 as $\alpha$ approaches $\alpha_u$ from below and there is no convective steady state for $\alpha > \alpha_u$ (see also figure \ref{fig:Nu-contours}).  We expect that data at higher $Ra$ would reduce the best-fit $\alpha_2$ to be at or below $\alpha_2 \sim Ra^{0.25}$.

\begin{figure}
   \centering
   \includegraphics[width=\textwidth]{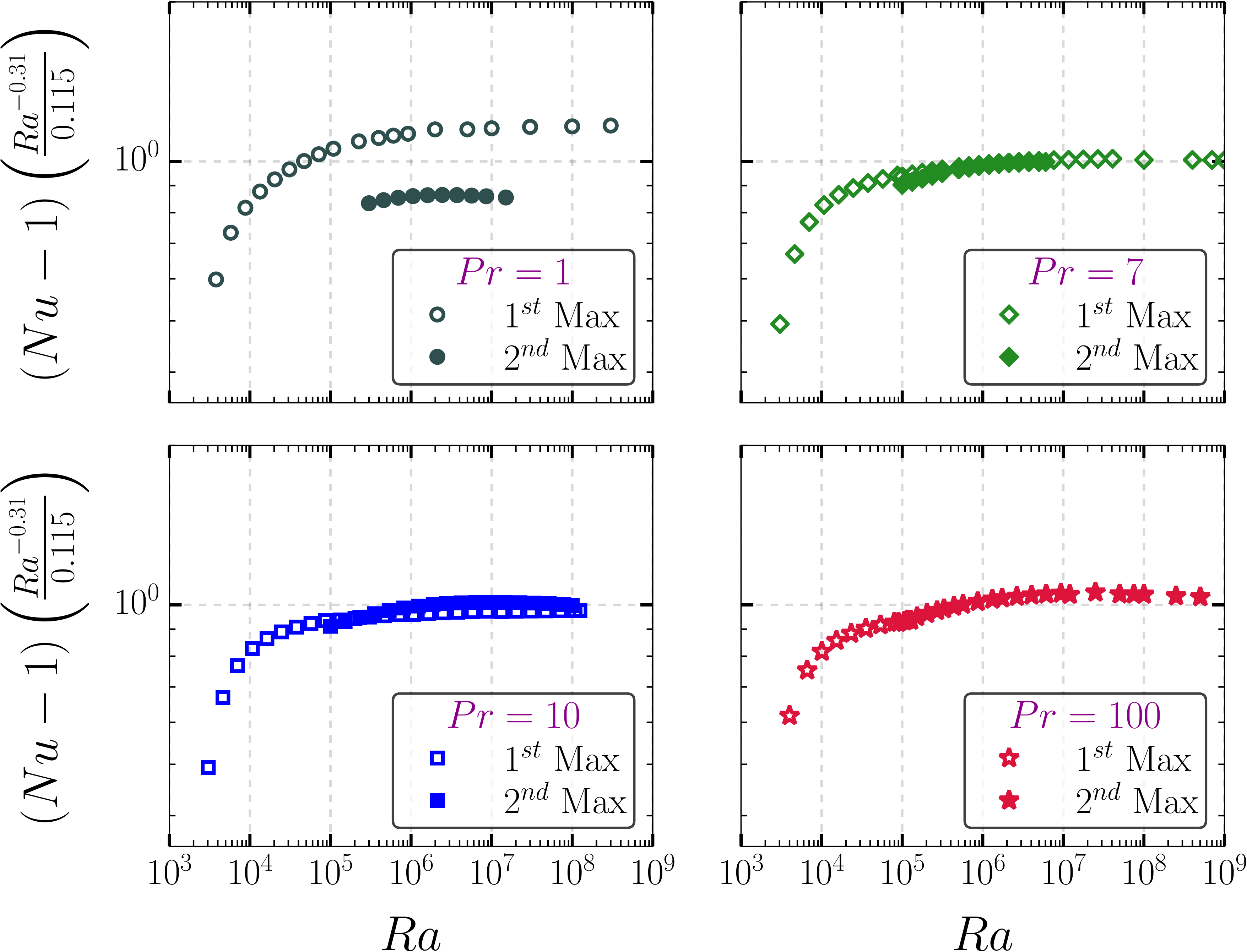}
   \caption{$Nu$ compensated by $0.115\, Ra^{0.31}$ for various $Pr$.  Open markers:  solutions from the first maximum.  Filled markers:  solutions from the second maximum.  Optimal solutions occur when one curve is above the other curve.}
   \label{fig:Nu-Ra-comp}
 \end{figure}

Figure~\ref{fig:Nu-Ra-comp} presents $Nu$ compensated by $0.115\, Ra^{0.31}$ which is the best least square fit scaling of $Nu$ at $Pr=7$ observed in figure~\ref{fig:Nu-Ra}.  In particular, we consider the $Nu$ vs. $Ra$ scaling at various $Pr$, and for each local maximum of $Nu$ (first and second local maxima).
We observe that the $Nu(Ra)$ scaling laws that result from tracking the first or second local maxima are not very different.  The most obvious discrepancy in $Nu$ between the first and second local maxima occurs for $Pr=1$.  However, the compensated plots are both flat over their respective $Ra$ ranges.  This indicates that the scaling exponent $\gamma=0.31$ is observed for both types of solutions at $Pr=1$.  The fact that the scaling of the optimal solution at $Pr=1$ is greater than unity can be attributed to a different prefactor in the scaling law (greater than $0.115$).  Similarly, the solution at the second local maximum has a prefactor less than $0.115$.  At $Pr=7$ and $Pr=10$ there is almost no discernible difference in the scaling laws between to the two types of solutions.  When $Pr=100$ the optimal solution also closely follows the scaling shown in figure~\ref{fig:Nu-Ra}.

\subsection{Coherent Structures and Optimal Heat Transport}\label{sec:CS}
Sections~\ref{sec:MM} and~\ref{sec:SL} showed that the $Nu$-maximizing wavenumbers are larger than the wavenumber $\alpha = 1.5585$ corresponding to the onset of convection (the primary wavenumber).  Thus the maximizing solutions have smaller horizontal scales than their non-optimal counterparts in the primary box. Here we compare the details of the optimal and non-optimal coherent structures.

\begin{figure}
  \centering
  \includegraphics[width=0.75\textwidth]{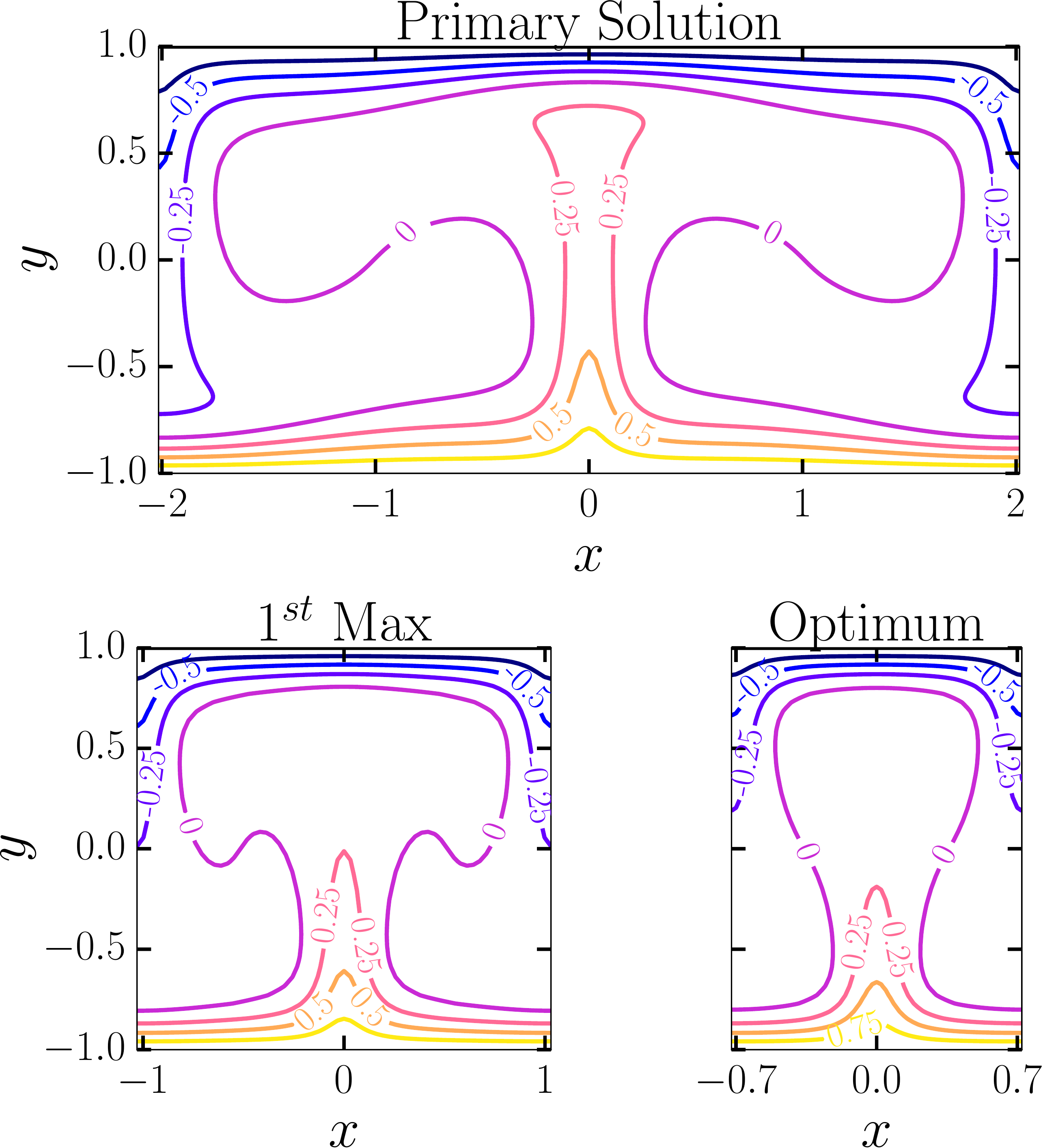}
  \caption{Temperature contours at $Ra=1\times 10^{5}$ and $Pr=100$.  The top figure presents the primary solution ($\alpha=1.5585$, $Nu=4.429$), the bottom left figure represents contours at the first local maximum ($\alpha=3.0$, $Nu=4.819$), and the bottom right figure represents contours at the second local maximum ($\alpha=4.3$, $Nu=4.821$).  Note that the second local maximum is the optimal solution at the $Ra$ considered.
  }
  \label{fig:TC-Pr100}
\end{figure}

\begin{figure}
  \centering
  \includegraphics[width=\textwidth]{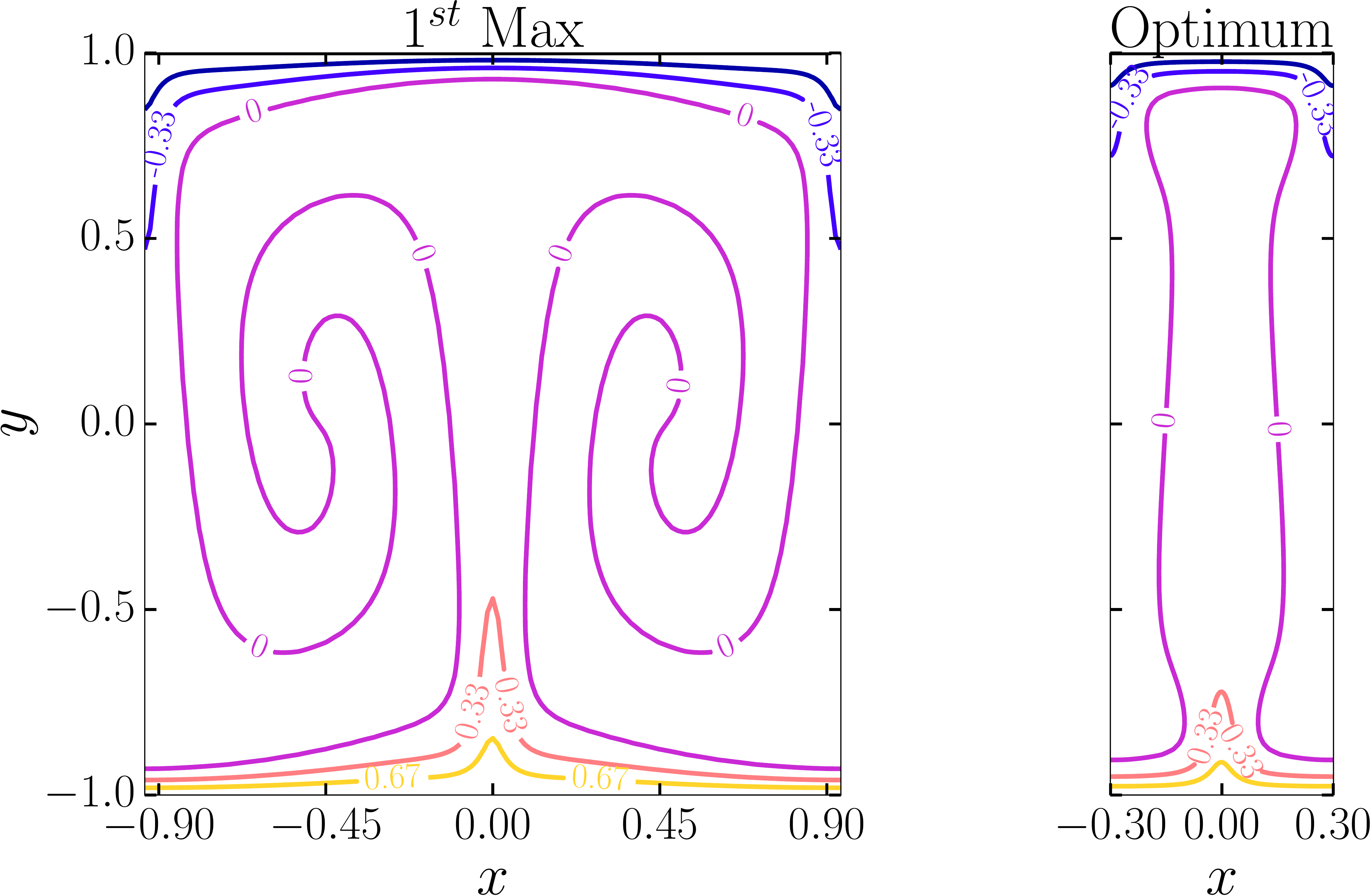}
  \caption{Temperature contours at $Pr=10$ and $Ra=2\times 10^{6}$ corresponding to optimal $Nu$ occur at $\alpha = 10.45$ and $Nu = 11.391$ (right plot).  Temperature contours for the first local maximum of $Nu$ occur at $\alpha = 3.35$ and $Nu=11.009$ (left plot). 
  }
  \label{fig:Pr10-contours}
\end{figure}

In figure~\ref{fig:TC-Pr100}, contours of temperature in the channel are plotted for $Pr = 100$ at $Ra = 1 \times 10^5$.  The top plot is the non-optimal temperature in the primary box with wavenumber $\alpha = 1.5585$.  The main features are the thick horizontal temperature arms emanating from the main plume that roll up with increasing $Ra$.  By contrast, the plumes in the bottom two plots maximize vertical heat transport and exhibit much less, if any,  horizontal structure. In fact, the thermal plume corresponding to the optimal solution (bottom right) does not have any spiral arms or lobes; the heat transport is almost exclusively straight up and down in the vertical direction.  Similar results are shown in figure~\ref{fig:Pr10-contours} for $Pr = 10$ at $Ra = 2\times 10^6$.  Again the optimal temperature structure is a single plume without significant horizontal structure (right; second local maximum), while the non-optimal plume (left; first local maximum) has coiling arms emanating from the central plume.  Figures~\ref{fig:TC-Pr100} and~\ref{fig:Pr10-contours} show representative optimal solutions in fluids with $Pr > 7$ at large-enough $Ra$, for which the optimal solution occurs at small scales (large $\alpha$; the second maximum).

\begin{figure}
  \centering
  \includegraphics[width=\textwidth]{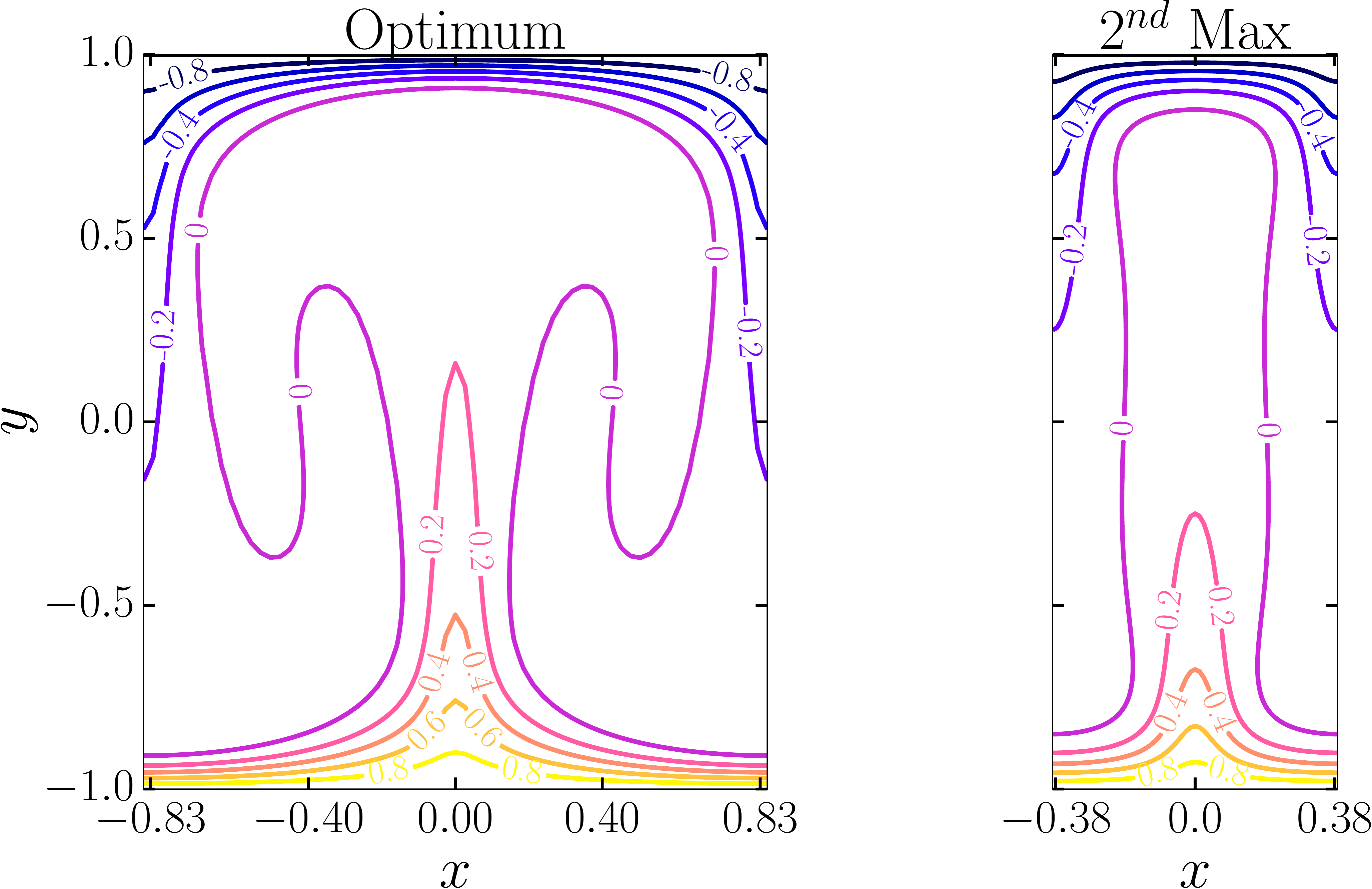}
  \caption{Temperature contours at $Pr=1$ and $Ra=3\times 10^{5}$ corresponding to optimal $Nu$ occur at $\alpha = 3.7$ and $Nu=7.058$ (left plot).  Temperature contours for the second local maximum of $Nu$ (right plot) occur at $\alpha = 8.1$ and $Nu=5.473$.}
  \label{fig:Pr1-contours}
\end{figure}

However, as shown in figure~\ref{fig:Pr1-contours}, the temperature structure of optimal solutions for $Pr \lesssim 7$ seems to contradict the simple optimal structure seen for $Pr > 7$.  The figure~\ref{fig:Pr1-contours} contours are for $Pr=1$ at $Ra = 5 \times 10^5$, with optimal solution on the left ($\alpha = 3.7$; first maximum), and non-optimal solution on the right ($\alpha = 8.1$; second maximum).  As expected, horizontal structure is prohibited by the large-$\alpha$ maximum, but in this case there is relatively weak vertical heat transport associated with the central plume. The smaller wavenumber (larger horizontal scale) allows for downdrafts to the left and right of center, which in turn allows for more vigorous updraft of heat in the center, and thus an overall larger vertical heat transport. Contours of the optimal vertical velocity (not shown here) confirm that the vertical velocity is largest in the middle of the channel.  

Thus, depending on the $Pr$ and $Ra$ numbers, there is a delicate balance determining whether or not the optimal solution will have horizontal structure in the form of coiling arms emanating from the central plume. For $Pr > 7$ at large-enough $Ra$, the optimal solution is the `second maximum' corresponding to small horizontal scale that does not admit significant horizontal structure. In this case, top to bottom heat transport through the central plume is the most efficient conduit for overall vertical heat transport. For $Pr \lesssim 7$, the optimal solution is the `first maximum' corresponding to a larger horizontal scale, wherein the largest overall vertical heat transport is achieved through a vigorous central updraft accompanied by adjacent downdrafts.

\begin{figure}
  \centering
    \includegraphics[height=0.4\textheight]{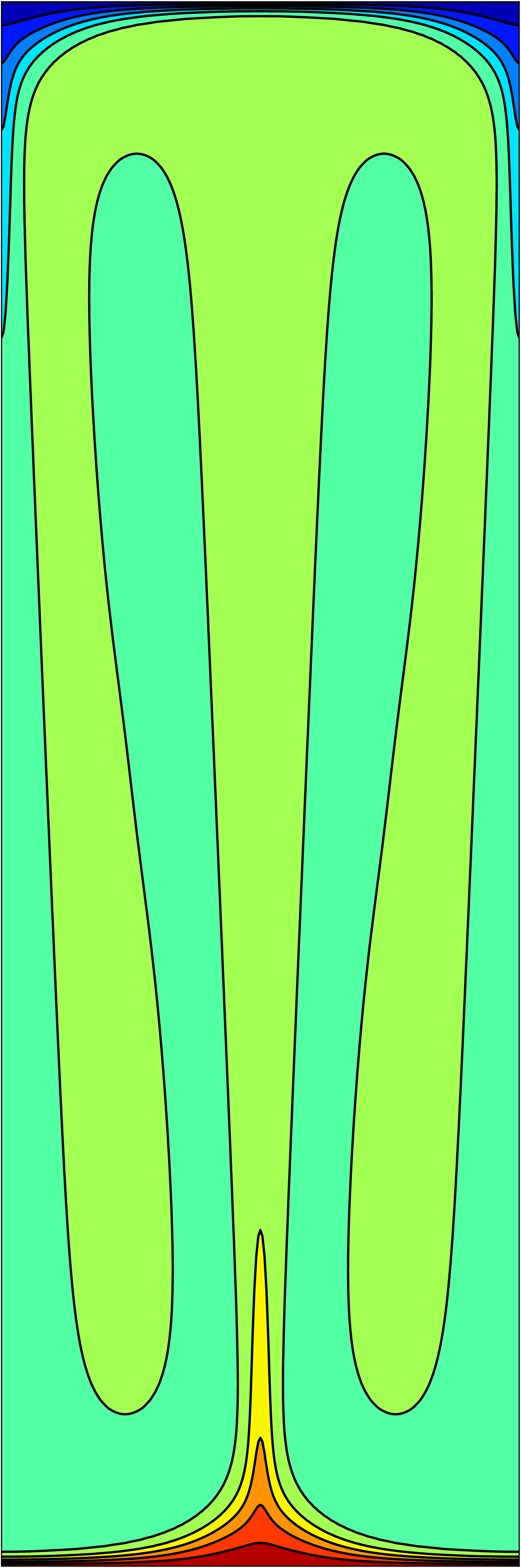}
    \includegraphics[height=0.4\textheight]{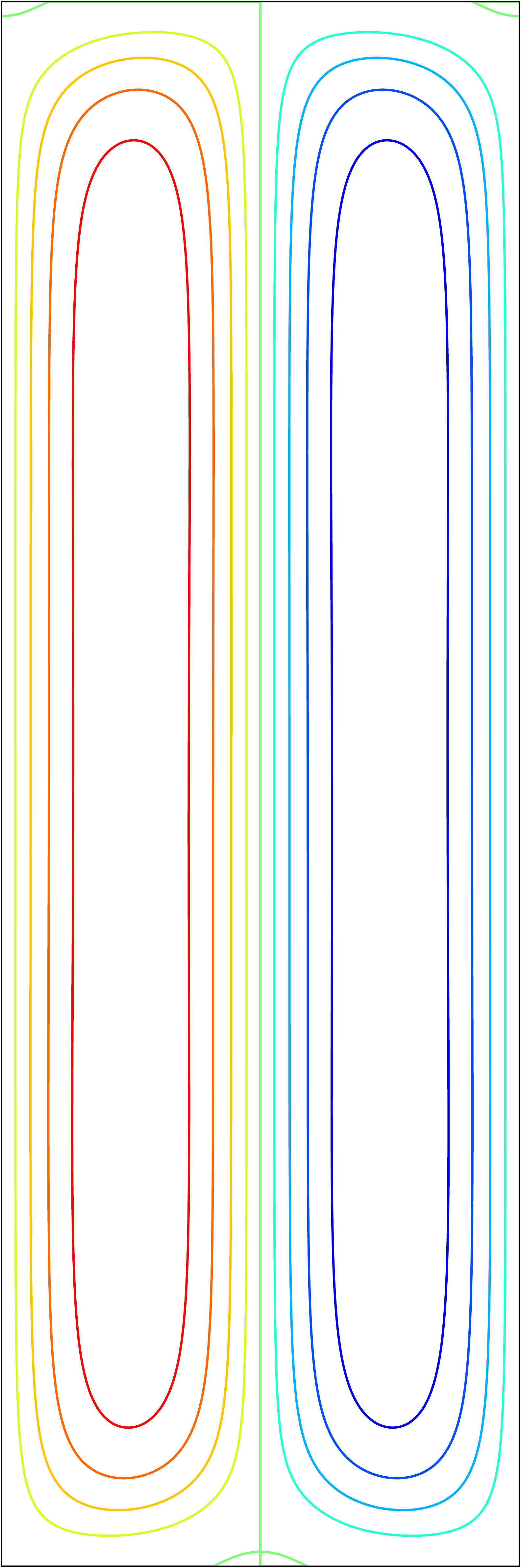}
    \qquad \qquad
    \includegraphics[height=0.4\textheight]{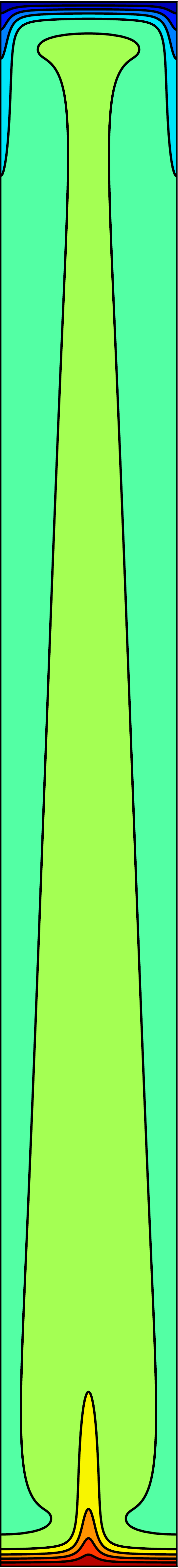}
    \includegraphics[height=0.4\textheight]{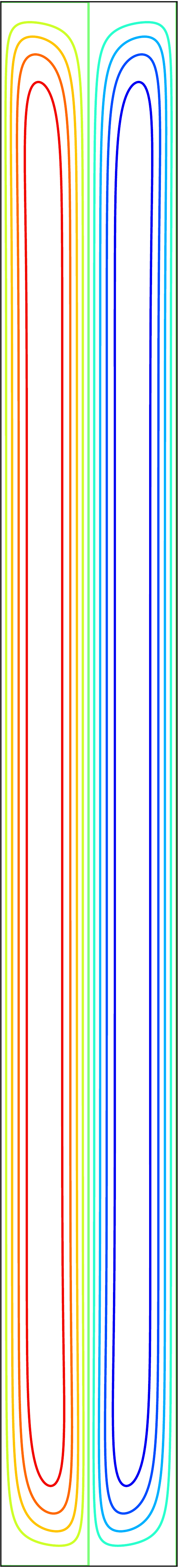}
  \caption{$Ra=10^8$. Two leftmost plots are temperature contours and streamlines for the optimal solution at $Pr=1$ with $\alpha=9.5$ and $Nu=41.49$.  Two rightmost plots are temperature contours and streamlines for the optimal solution at $Pr=10$ with $\alpha=27.86$ and $Nu=35.63$.  As usual $y\in\left[-1, 1\right]$ and $x\in\left[-\pi/\alpha, \pi/\alpha\right]$.
}
  \label{fig:Optimal-solution-HighRa}
\end{figure}

\begin{figure}
  \centering
  \includegraphics[width=0.7\textwidth]{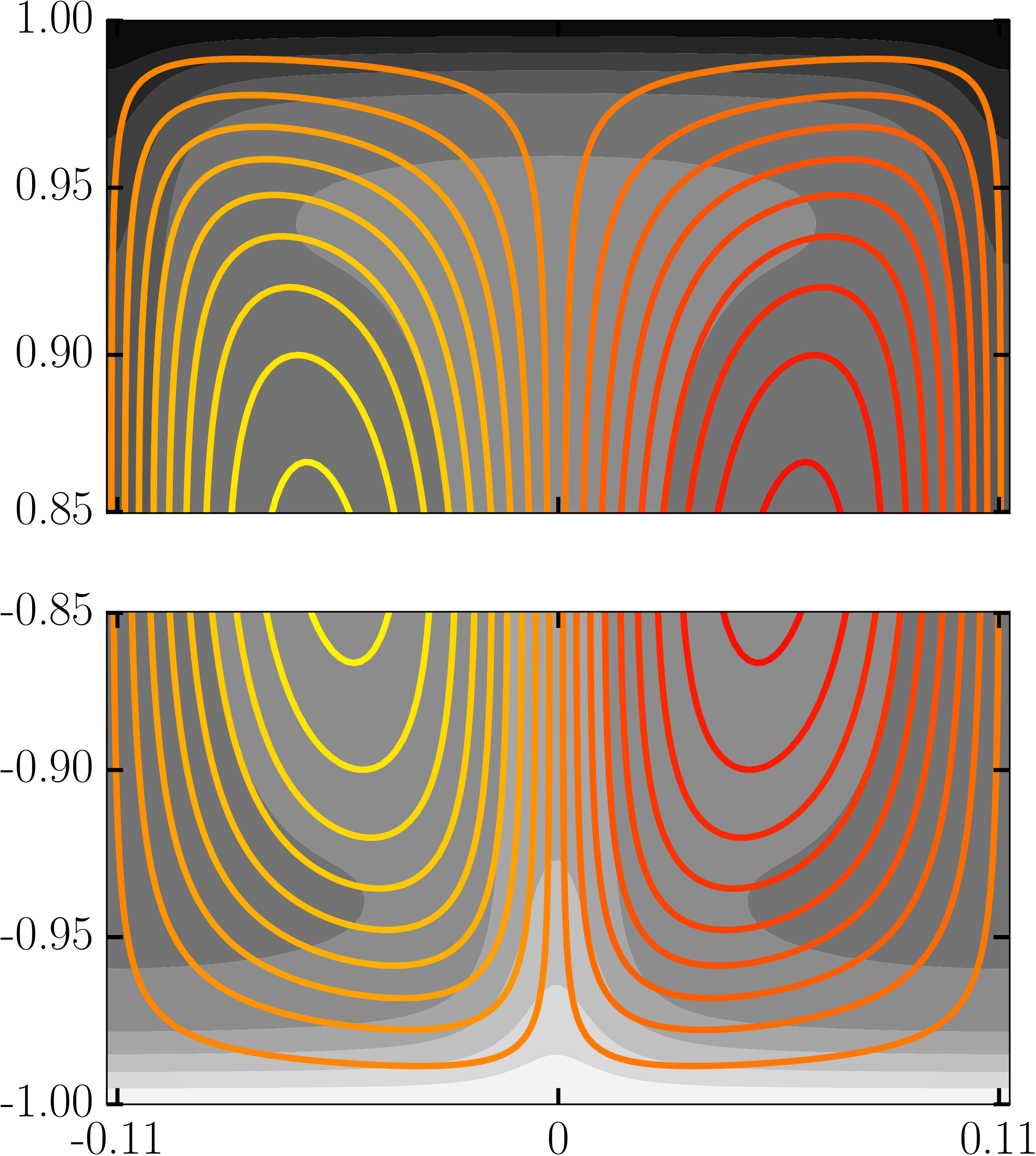}
  \caption{Streamlines of the optimal solution at $Pr=10$ and $Ra=1\times 10^{8}$ near the top (top plot) and bottom (bottom plot) superposed on contours of temperature where $x\in\left[-\pi/\alpha, \pi/\alpha\right]$ and $\alpha=27.86$.} 
  \label{fig:zoomed-streamlines-Pr10-HighRa}
\end{figure}

We close this subsection with a visualization of the optimizing solution at $Ra = 10^{8}$ for $Pr=1$ and $Pr=10$. (figures~\ref{fig:Optimal-solution-HighRa} and~\ref{fig:zoomed-streamlines-Pr10-HighRa}).  The two leftmost plots in figure~\ref{fig:Optimal-solution-HighRa} present optimal temperature contours and corresponding streamlines at $Pr=1$.  The rightmost plots show optimal temperature contours and corresponding streamlines at $Pr=10$.
Figure~\ref{fig:zoomed-streamlines-Pr10-HighRa} shows the top (top plot) and bottom (bottom plot) portions of the optimal temperature field with streamlines superposed at $Pr=10$.  A main feature to note is that the streamlines deeply penetrate the boundary layer.
We remark that~\citet{hassanzadeh2014wall} found optimal solutions for a related problem in which the class of solutions was expanded to include all divergence free velocity fields with fixed enstrophy and free-slip boundary conditions.  The aspect ratio in that work was found to scale as $Ra^{1/4}$ as in the present work.  One notable difference is that we do not observe the formation of a recirculation region near the top of the channel.

\subsection{Mean Temperature Profiles and $Nu\lr{Pr}$}\label{sec:MTP}
\begin{figure}
  \centering
  \includegraphics[width=0.85\textwidth]{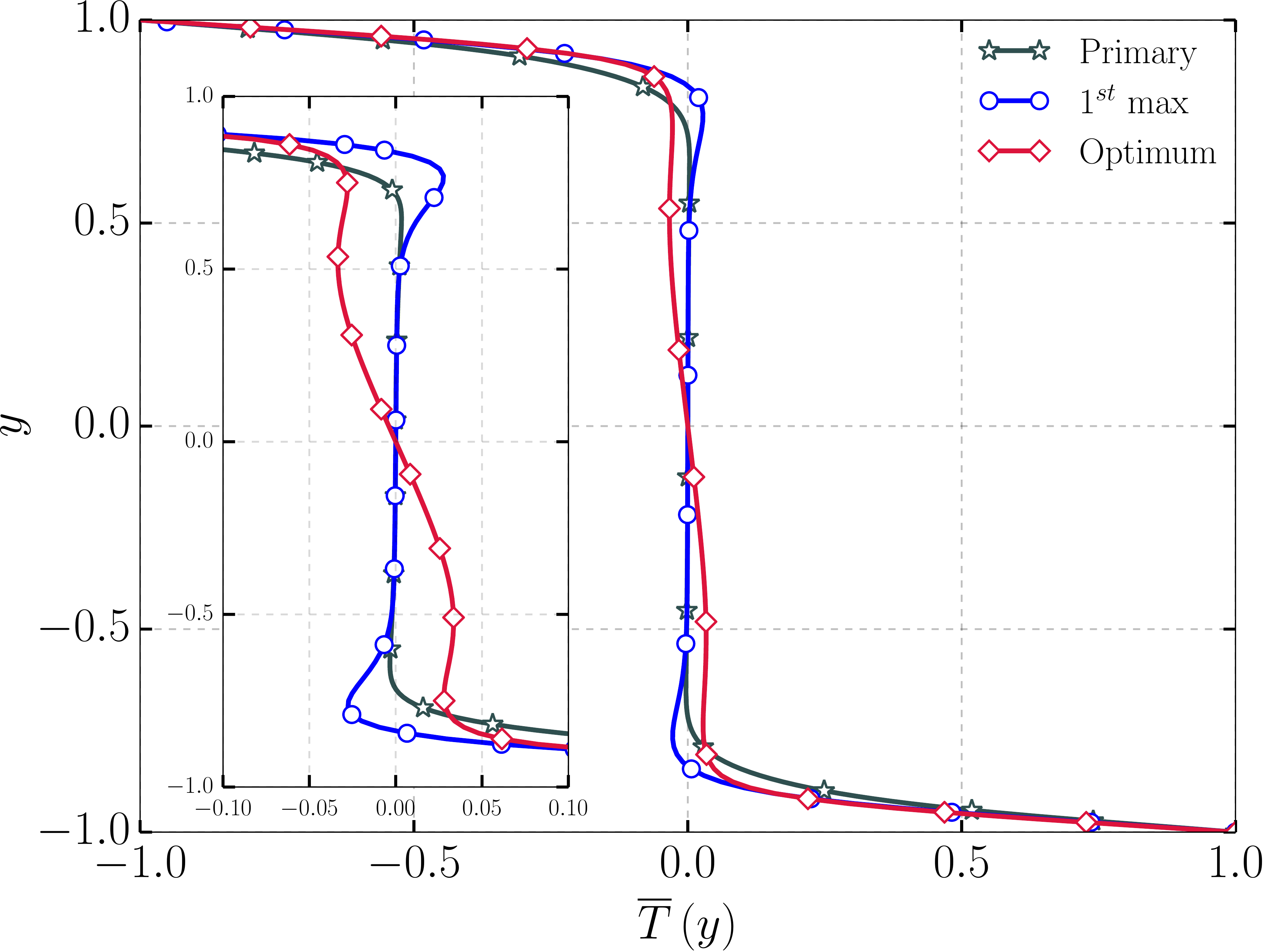}
  \caption{Horizontally-averaged temperature profiles for $Pr=10$ at $Ra=2\times 10^{6}$.}
  \label{fig:Pr10}
\end{figure}

\begin{figure}
  \centering
  \includegraphics[width=0.85\textwidth]{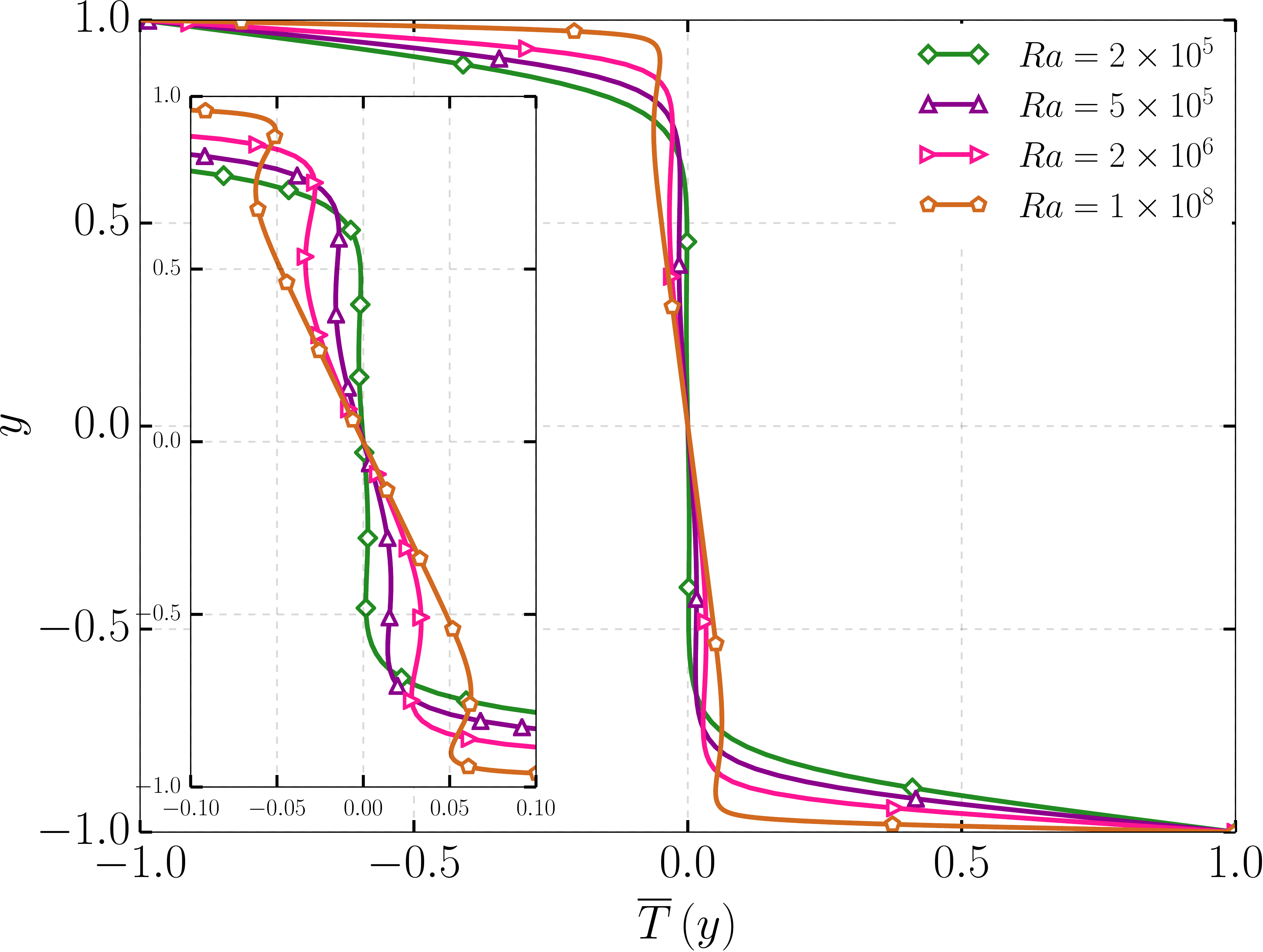}
  \caption{Horizontally-averaged optimal temperature profiles at $Pr=10$ for $2\times 10^{5} \leq Ra\leq 1\times 10^{8}$.}
  \label{fig:Tave-Pr10Ra}
\end{figure}

We now consider mean temperature profiles of the optimal solutions introduced in section~\ref{sec:CS}.  Figures~\ref{fig:Pr10} and~\ref{fig:Tave-Pr10Ra} present mean temperature profiles at $Pr=10$.  First, in figure~\ref{fig:Pr10} we compare mean temperature profiles at $Ra=2\times 10^{6}$ of the primary solution, the optimal solution, and the solution that gives the first maximum.  The primary solution is nearly marginally stable, possessing a nearly isothermal region though the core of the channel.  However, a very slight stable stratification of the primary solution can be observed in the inset of figure~\ref{fig:Pr10} just above $y=0.5$.  The solution corresponding to the first maximum is marginally stable in the center of the channel but exhibits stably stratified transition regions between the center and the unstably stratified boundary layers.  The optimal solution, on the other hand, is unstably stratified in the center of the channel, but maintains the stably stratified transition regions and the unstable boundary layers.  We note that non-monotonic temperature profiles have been considered in the infinite $Pr$ case~\citep{doering2006bounds}.  Interestingly, each solution presented in figure~\ref{fig:Pr10} is close to the marginally stable configuration constructed by~\citet{malkus1954heat} and resulting in the prediction of $Nu\sim Ra^{1/3}$. 


As shown in figure~\ref{fig:Nu-Ra} the scaling of the optimal solution $Nu\sim Ra^{0.31}$ is less than $Nu \sim Ra^{1/3}$ which might be attributed to small departures from marginal stability.  Figure~\ref{fig:Tave-Pr10Ra} shows optimal mean temperature profiles at $Pr=10$ for $2\times 10^{5}\leq Ra \leq 1\times 10^{8}$.  Each of these profiles exhibits unstable stratification in the bulk and in the boundary layers, but still possesses the stably stratified transition regions.  Perhaps surprisingly, the optimal solutions become more unstably stratified in the bulk as $Ra$ increases, and thus departure from marginal stability seems to grow.  The latter is related to the single updraft nature of the optimal solutions, which have increasing horizontal wavenumber and diminishing horizontal structure; there are no significant horizontal arms or coils to bring colder fluid beneath warmer fluid.

\begin{figure}
  \centering
  \centering
  \includegraphics[width=0.85\textwidth]{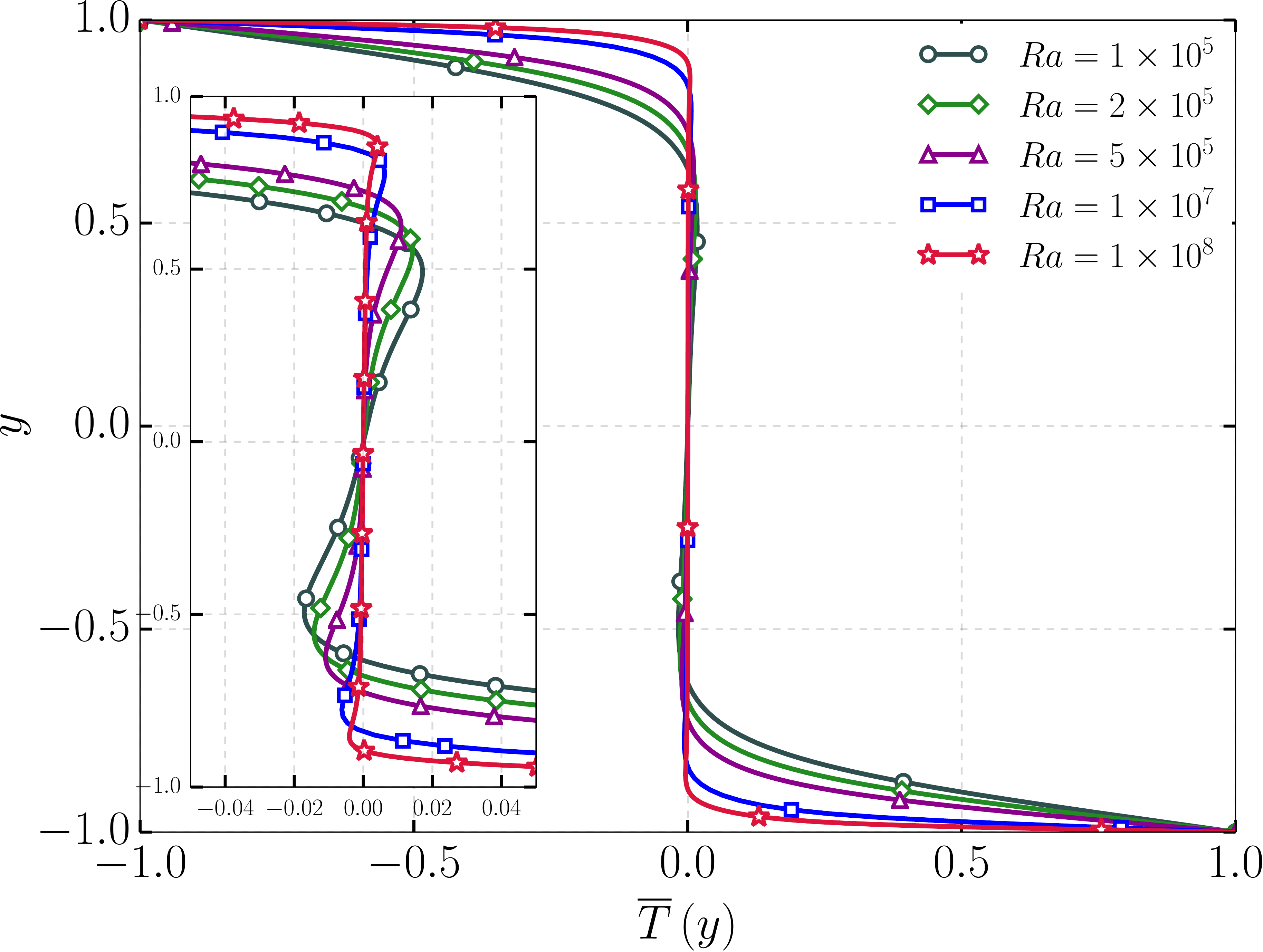}
  \caption{Horizontally-averaged optimal temperature profiles at $Pr=1$ for $10^{5} \leq Ra\leq 10^{8}$.}
  \label{fig:Tave-Pr1}
\end{figure}

A different trend is observed for $Pr \lesssim 7$, as illustrated in figure~\ref{fig:Tave-Pr1}, showing horizontally averaged optimal temperature profiles at $Pr=1$ for $10^{5}\leq Ra \leq 10^{8}$.  At first glance, these profiles all appear to be marginally stable.  The inset in figure~\ref{fig:Tave-Pr1} shows that this observation is not quite the entire story.  Indeed, near the center of the channel the profiles are nearly marginally stable, with a tendency toward stable stratification, especially at lower $Ra$.  Stably stratified transition regions between the center of the channel and the boundary layers are once again present.  The mean stable stratification is linked to the two downdrafts on either side of the central updraft, which bring colder fluid underneath warmer fluid in a stable configuration.  

\begin{figure}
  \centering
  \includegraphics[width=0.85\textwidth]{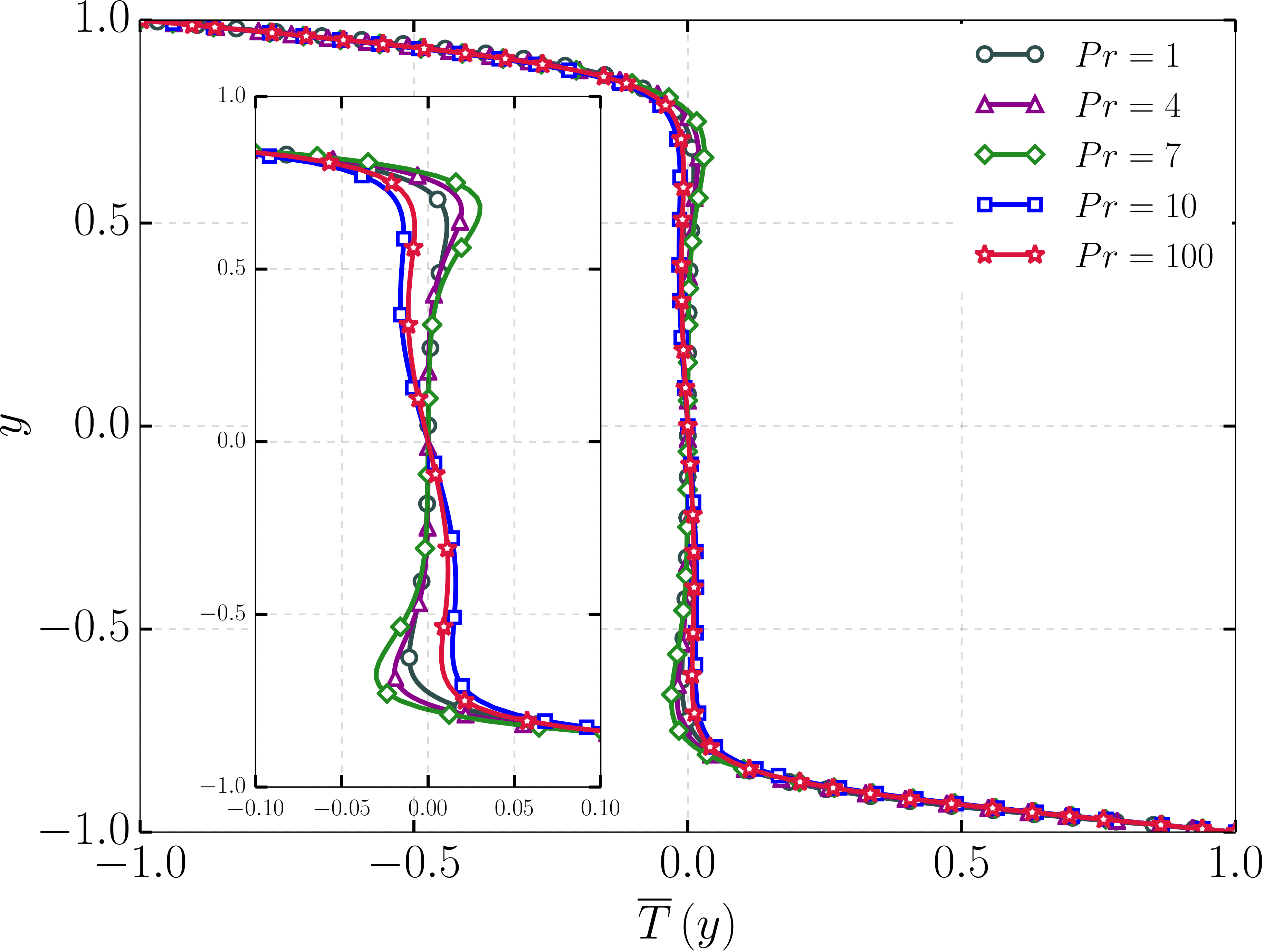}
  \caption{Horizontally-averaged optimal temperature profiles at $Ra=5\times 10^{5}$ for $Pr=1$, $Pr=4$, $Pr=7$, $Pr=10$ and $Pr=100$.}
  \label{fig:Tave-Pr}
\end{figure}

Figures~\ref{fig:Pr10},~\ref{fig:Tave-Pr10Ra} and~\ref{fig:Tave-Pr1} suggest that for $Pr\lesssim 7$ optimal solutions have a stably stratified bulk region whereas the optimal solutions for $Pr>7$ have an unstably stratified bulk region.  Figure~\ref{fig:Tave-Pr} provides more evidence of this emerging picture, presenting optimal solutions at $Ra=5\times 10^{5}$ for $Pr = 1, 4, 7, 10, 100$.  Indeed, the optimal mean temperature profiles for $Pr = 1, 4, 7$ are stably stratified near the middle of the channel, while those for $Pr=10$ and $Pr=100$ are unstably stratified.  The stably stratified transition regions between the middle of the channel and the boundary layers are present at all $Pr$.  We summarize the three regions observed in the optimal temperature profiles as follows:
\begin{enumerate}
  \item The temperature boundary conditions produce unstable boundary layers.
  \item The central region of the channel may be stably stratified ($Pr\lesssim 7$) or unstably stratified ($Pr>7$).
  \item The transition regions between the channel center and the boundary layers were always observed to be stably stratified in the optimal transport solutions.
\end{enumerate}

\begin{figure}
  \centering
  \includegraphics[width=0.85\textwidth]{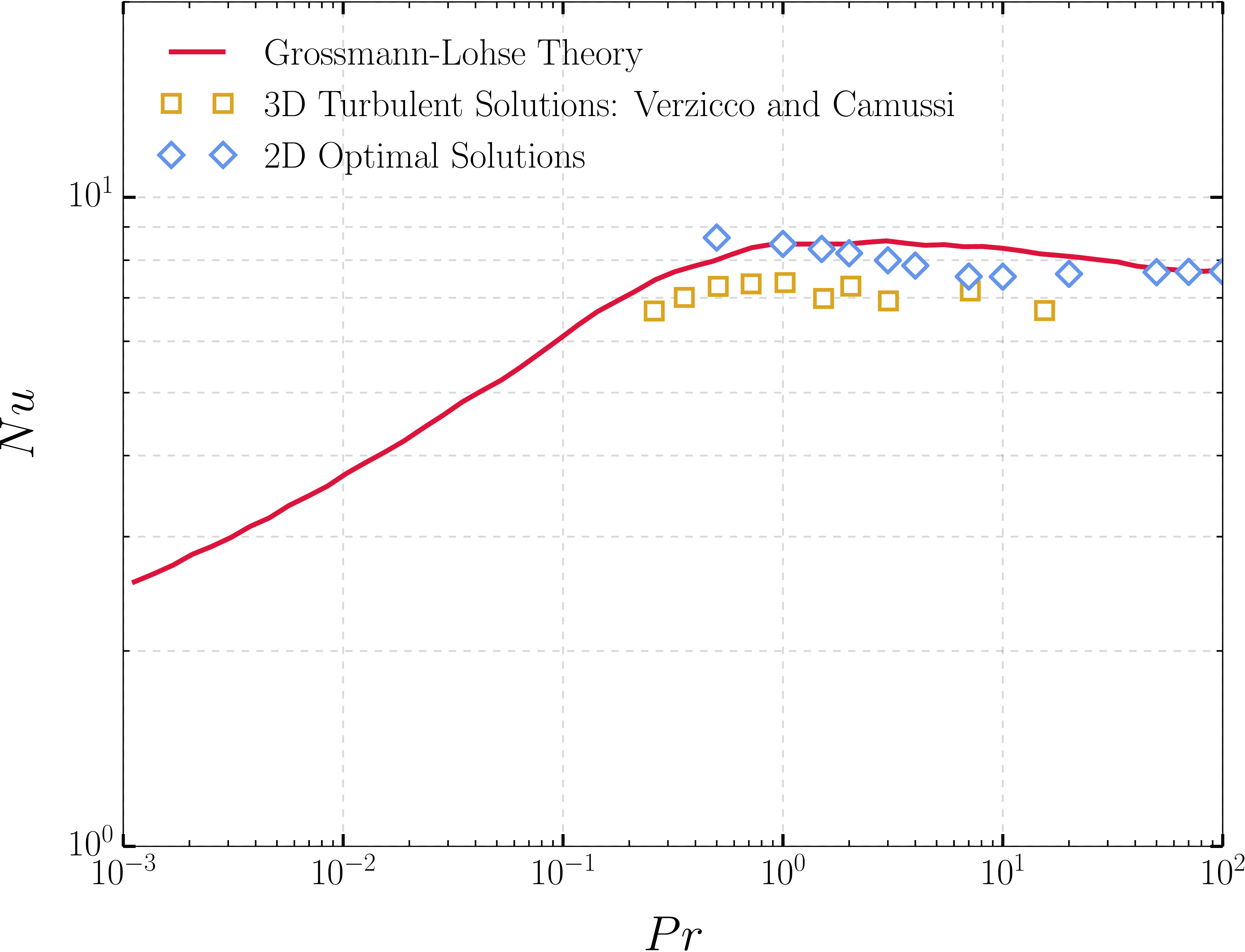}
  \caption{Comparison of 2D optimal solutions to the Grossmann-Lohse theory and to 3D turbulent solutions~\citep{verzicco1999prandtl} for $0.25\leq Pr \leq 100$ at $Ra = 5\times 10^{5}$.}
  \label{fig:Tave-Pr-compare}
\end{figure}

Finally, we compare the two dimensional optimal solutions found in the present work to the Grossmann-Lohse (GL) theory of turbulent convection~\citep{grossmann2000scaling} and to fully three-dimensional numerical simulations~\citep{verzicco1999prandtl}.  In figure~\ref{fig:Tave-Pr-compare} we show the variation of $Nu$ with $Pr$ for $0.5 \leq Pr \leq 100$ at $Ra = 5\times 10^{5}$.  The work presented here corresponds to regions $\rom{1}_{u}$ and $\rom{2}_{u}$ as well as the beginning of region $\rom{4}_{u}$ in the $Pr-Ra$ phase diagram of the GL theory~\citep{grossmann2001thermal}.  In each of these regions the thermal boundary layer is nested within the viscous boundary layer.
We remark that the 2D optimal solutions are in agreement with both the GL theory as well as the 3D turbulent solutions.  This intriguing observation begs the question: do the 2D steady solutions computed in the present work play a role in the dynamics of 3D turbulent convection?  For future consideration of the possible connection, one must consider aspect-ratio effects as well as the difference in dimensionality (2D vs.\ 3D).  The theoretical \RB problem has an essentially infinite aspect ratio, whereas the aspect-ratio dependent parameters in the GL theory have been fit to experiments with aspect ratio $\Gamma\leq 1$. The turbulent simulations of~\citet{verzicco1999prandtl} were performed in a cylindrical cell with $\Gamma = 1$.  There is evidence, however, that aspect ratio dependence may be weak~\citep{ahlers2009heat}.  On the other hand, the sensitivity of $Nu$ to the large scale circulation arising in finite aspect ratio configurations remains an open issue~\citep{ahlers2009heat}.
\section{Conclusions}\label{sec:conclusions}
Optimal transport solutions for 2D \RB convection have been computed for $Ra$ up to $10^{9}$ for $Pr\geq 1$.  This work is an extension to that performed by~\citet{waleffe2015heat} at $Pr=7$.  Most of the solutions in the present work have been found using a \flowmap algorithm in conjunction with a Jacobian-free Newton-Krylov solver to find steady solutions to the \OB equations. Some of the high $Ra$, $Pr=1$ and $Pr=7$ solutions were computed with code 2 in \citet{waleffe2015heat}.  An optimal solution at a given $Ra$ and $Pr$ represents the steady solution that maximizes heat transport among all horizontal wavenumbers.  One goal is to make contact with rigorous upper bound theory~\citep{howard1963heat, busse1969howard, doering1996variational} by computing optimal solutions constrained exactly by the \OB equations.  Another intention is to compare scaling laws derived from experiments~\citep{niemela2000turbulent,he2012heat} and computational studies~\citep{amati2005turbulent} of turbulent \RB convection with our optimal solutions in the spirit of Malkus's original idea that turbulence maximizes heat transport subject to some marginal stability constraints~\citep{malkus1954heat}.  A significant portion of this work is aimed at exploring the effect of $Pr$ on scaling laws and optimal solutions.

We have considered $Pr=1, 4, 7, 10, 100$ and our results indicate that there is little to no variation of scaling of $Nu$ with $Pr$ (for the parameter ranges considered).  This is consistent with studies in turbulent \RB convection which report that $Nu\lr{Pr}$ is relatively constant for $Pr\gtrsim 1$~\citep[see][figure 5]{ahlers2009heat}.  The optimal solutions exhibit a scaling of the form $Nu\sim Ra^{0.31}$ which is in agreement with that found by~\citet{waleffe2015heat}.  Moreover, horizontally averaged profiles of the temperature field are close to a marginally stable profile, roughly consistent with the prediction by~\citet{malkus1954heat} leading to $Nu\sim Ra^{1/3}$.  However, the lack of $Pr$ dependence for scaling relations is not the entire story.  Indeed, $Pr$ is shown to have a significant impact on the scale of the optimal solution.  When considering the variation of $Nu$ with $\alpha$, we observe that, above a certain $Ra$, two local maxima exist; one at a smaller scale and the other at a larger scale.  There are two regions partitioned by $Pr\approx 7$:  for $Pr>7$ optimal transport solutions occur at smaller scales and for $Pr\lesssim 7$ optimal transport solutions occur at larger scales.

One might expect optimal heat transport to be realized by direct `wall-to-wall' transport of hot and cold fluid in narrow plumes with little horizontal structure and no returning downdrafts of hot fluid and updrafts of cold fluid.  Inspection of the optimizing solutions reveals that this structure is associated with the `second maximum' at small scales, and optimizes heat transport for $Pr > 7$ at large-enough $Ra$.  For $Pr \lesssim 7$, the optimal heat transport occurs at the `first maximum', and is associated with a larger horizontal scale. In this case, the vigorous central updraft is accompanied by adjacent left/right downdrafts, leading to a slightly stable interior mean temperature profile. Evidently, this is another way to minimize the average thermal boundary layer to maximize transport.  Alternatively, with free-slip boundary conditions and an expanded class of velocity fields, optimal transport solutions were found to exhibit recirculation near the top of the channel~\citep{hassanzadeh2014wall}.  Thus there is a delicate balance when optimizing over all possible velocity fields:  the velocity must be strong enough to efficiently transport heat vertically, but not so strong that downdrafts of heat overcompensate for the vertical transport. It is unclear why $Pr\approx 7$ should be the demarcation between smaller-scale optima consisting of a central updraft only, and larger-scale optima admitting some horizontal structure (downdrafts adjacent to the central updraft).

An obvious extension of the present work is to consider a wider range of $Pr$ and $Ra$ numbers and in particular $Pr<1$.  Stricter spatial and temporal resolutions will be necessary, requiring algorithmic modifications including possibly code parallelization.  Another direction of our future work is understanding the connections between the optimal steady solutions and turbulent flows. It has been conjectured that such unstable solutions actually control the dynamics of turbulent flows~\citep{waleffe2009exact} and the close agreement between the $Nu(Ra)$ data for optimum 2D solutions and 3D turbulent data is most intriguing.  Is it possible to detect the signature of the optimal solutions within the turbulence data?  One step in this direction involves performing a fully resolved simulation at high $Ra$ and collecting local snapshots of the fields at times and locations where the Nusselt number achieves its maximal values in order to `educe' a characteristic optimum transport structure~\citep{stretch1990automated}.  These snapshots can be compared to the  coherent structures educed through other flow decompositions~\citep[e.g.,][]{smith2005low,schmid2010dynamic}.  Thus one may begin to explore statistics such as how often the optimal structures are sampled by the turbulent flow. More sophisticated analyses of the state space and the role of the stable and unstable manifolds of the optimum solutions should be easier to perform than in 3D shear flows~\citep{gibson2008visualizing}.  We hope that such investigations will be the subject of forthcoming publications.

\section{Acknolwedgments}
The authors would like to thank Charlie Doering (University of Michigan) for insightful discussions and suggestions.  We are also grateful for discussions with David Goluskin (University of Michigan), Andre Souza (University of Michigan), Greg Chini (University of New Hampshire) and Anakewit Boonkasame (University of Wisconsin-Madison).  This work was partially supported by DMS-1147523 (DS), DMS-0807349 (FW) and DMS-1008396 (LMS).

\printbibliography

\appendix

\section{Boundary Conditions and Time Integration}\label{app:BCTI}

\subsection{Time Integration}\label{app:TI}

We now apply the (3,4,3) IMEX-RK method developed by~\citet{ascher1995implicit} to the \RB problem governed by equations~\eqref{eq:phi-eq} and~\eqref{eq:energy-nondim}.  We rewrite the equations concisely as
\begin{align}
  \pdeone{\phi}{t} + \pdeone{}{x}\mathcal{N}_{\phi} &= \nund\nabla^{2}\phi + \pden{T}{x}{2} \\
  \pdeone{T}{t} + \mathcal{N}_{T} &= \kand\nabla^{2}T
\end{align}
where
\begin{align}
  \mathcal{N}_{\phi} = \mathcal{N}_{\phi}\lr{\phi, u, v} &= u\phi - v\nabla^{2}u \\
  \mathcal{N}_{T}    = \mathcal{N}_{T}\lr{T, u, v}       &= \mathbf{u}\cdot\nabla T.
\end{align}
We will treat the diffusion terms implicitly and all other terms explicitly.  We denote by $\phi^{n}$ the field at the current time step.  To evolve $\phi$ to step $n+1$ we have,
\begin{align}
  \phi^{n+1} = \phi^{n} + \Delta t \sum_{j=1}^{3}b^{j}\lr{K_{\phi}^{j} + \widehat{K}_{\phi}^{j+1}}
\end{align}
where
\begin{align}
  K_{\phi}^{j}           &=  \nund\nabla^{2}\phi^{i} \\
  \widehat{K}_{\phi}^{1} &= -\pdeone{}{x}\mathcal{N}_{\phi}^{n-1} + \pden{}{x}{2}T^{n-1} \\
  \widehat{K}_{\phi}^{i} &= -\pdeone{}{x}\mathcal{N}_{\phi}^{i-1} + \pden{}{x}{2}T^{i-1}, \quad i\neq 1.
\end{align}
At stage $i$ $\phi$ is given by
\begin{align}
  \lr{1 - a^{ii}\Delta t\nund\nabla^{2}}\phi^{i} = \phi^{n} + \Delta t\left[\sum_{j=1}^{i-1}{\lr{a^{ij}K_{\phi}^{j} + \widehat{a}^{i+1,j+1}\widehat{K}_{\phi}^{j+1}}} + \widehat{a}^{i+1,1}\widehat{K}_{\phi}^{1}\right]. \label{eq:phii}
\end{align}
The implicit operator is given by,
\begin{align}
  \mathcal{I}^{i}_{\phi} = 1 - a^{ii}\Delta t\nund\nabla^{2}.
\end{align}
The right hand side of equation~\eqref{eq:phii} represents $F_{\phi}^{i}$ in equation~\eqref{eq:phi-time-int} in appendix~\ref{app:BCs}.

The temperature at step $n+1$ is given by,
\begin{align}
  T^{n+1} = T^{n} + \Delta t \sum_{j=1}^{3}b^{j}\lr{K_{T}^{j} + \widehat{K}_{T}^{j+1}}
\end{align}
and
\begin{align}
  K_{T}^{j}           &=  \kand\nabla^{2}T^{i} \\
  \widehat{K}_{T}^{i} &= -\mathcal{N}_{T}^{i-1} , \quad i\neq 1 \\
  \widehat{K}_{T}^{1} &= -\mathcal{N}_{T}^{n-1}.
\end{align}
The temperature at stage $i$ is solved from the implicit equation
\begin{align}
  \lr{1-a^{ii}\Delta t\kand\nabla^{2}}T^{i} = T^{n} + \Delta t\left[\sum_{j=1}^{i-1}{\lr{a^{ij}K_{T}^{j} + \widehat{a}^{i+1,j+1}\widehat{K}_{T}^{j+1}}} + \widehat{a}^{i+1,1}\widehat{K}_{T}^{1}\right].
\end{align}
The values of $a^{ij}$, $\widehat{a}^{ij}$ and $b^{i}$ can be found in~\citet{ascher1995implicit}.

\subsection{Boundary Conditions}\label{app:BCs}

In order to satisfy the boundary conditions at stage $i$ we let
\begin{align}
  v^{i} = v_{p}^{i} + c_{1}v_{1}^{i} + c_{2}v_{2}^{i}
\end{align}
and solve three separate problems for $v_{p}^{i}$, $v_{1}^{i}$, and $v_{2}^{i}$ at each stage of the time integration where $v_{p}^{i}$ is the particular solution and $v_{1}$ and $v_{2}$ are two homogeneous solutions.  The three problems are:
\begin{subequations}
  \begin{equation}
    \begin{split}
      \mathcal{I}^{i}_{\phi}\phi_{p}^{i} &= F^{i}_{\phi}, \quad \phi_{p}^{i}\lr{\pm 1} = 0 \label{eq:phi-time-int} \\
              \nabla^{2}v_{p}^{i} &= \phi_{p}^{i}, \quad v_{p}^{i}\lr{\pm 1} = 0.
    \end{split}
  \end{equation}
  \begin{equation}
    \begin{split}
      \mathcal{I}^{i}\phi_{1}^{i} = 0, &\quad \phi_{1}^{i}\lr{1} = 0, \quad \phi_{1}^{i}\lr{-1} = 1 \\
      \nabla^{2}v_{1}^{i} = \phi_{1}^{i}, &\quad v_{1}^{i}\lr{\pm 1} = 0.
    \end{split}
  \end{equation}
  \begin{equation}
    \begin{split}
      \mathcal{I}^{i}\phi_{2}^{i} = 0, &\quad \phi_{2}^{i}\lr{1} = 1, \quad \phi_{2}^{i}\lr{-1} = 0 \\
      \nabla^{2}v_{2}^{i} = \phi_{2}^{i}, &\quad v_{2}^{i}\lr{\pm 1} = 0.
    \end{split}
  \end{equation}
\end{subequations}
$\mathcal{I}^{i}_{\phi}$ is the implicit time integration operator at stage $i$ and $F^{i}_{\phi}$ contains contributions from the explicit portion of the time integration at stage $i$ (see appendix~\ref{app:TI}).  The constants $c_{1}$ and $c_{2}$ are determined so that equation~\eqref{eq:noslip-nondim} for the boundary conditions is satisfied.  At each wall we have
\begin{align}
  0 = \pdeone{v^{i}}{y}\lr{\pm 1} = \pdeone{v_{p}^{i}}{y}\lr{\pm 1} + c_{1}\pdeone{v_{1}^{i}}{y}\lr{\pm 1} + c_{2}\pdeone{v_{2}^{i}}{y}\lr{\pm 1}.
\end{align}
This results in a $2\times 2$ matrix system that can be solved for $c_{1}$ and $c_{2}$,
\begin{align}
  \begin{bmatrix}
    \displaystyle \pdeone{v_{1}^{i}}{y}\lr{1}  & \displaystyle \pdeone{v_{2}^{i}}{y}\lr{1}  \\[0.75em]
    \displaystyle \pdeone{v_{1}^{i}}{y}\lr{-1} & \displaystyle \pdeone{v_{2}^{i}}{y}\lr{-1}
  \end{bmatrix}
  \begin{bmatrix}
    c_{1} \\[1.0em] c_{2}
  \end{bmatrix}
  =
  -
  \begin{bmatrix}
    \displaystyle \pdeone{v_{p}^{i}}{y}\lr{1} \\[0.75em]
    \displaystyle \pdeone{v_{p}^{i}}{y}\lr{-1}
  \end{bmatrix}
\end{align}
In the non-adaptive version of our code, $v_{1}^{i}$, $v_{2}^{i}$, $\phi_{1}^{i}$ and $\phi_{2}^{i}$ are computed as part of a preprocessing step and used in the determination of the constants.  When using adaptive time-step sizes $v_{1}^{i}$, $v_{2}^{i}$, $\phi_{1}^{i}$ and $\phi_{2}^{i}$ are computed just prior to the next time-integration.

\end{document}